\documentclass[reprint,aps,amsmath,amssymb,]{revtex4-2}
\usepackage{graphicx}
\usepackage{subcaption}
\usepackage{bm}
\usepackage{amsmath}

\begin{document}
\captionsetup[subfigure]{position=top,textfont=normalfont,singlelinecheck=off,justification=raggedright}


\title{Distinguishing spin pumping from spin rectification in organic semiconductor-based lateral spin pumping device architectures}


\author{Piotr Skalski}
\thanks{These two authors contributed equally}
\author{Olga Zadvorna}
\thanks{These two authors contributed equally}
\author{Deepak Venkateshvaran}
\author{Henning Sirringhaus}
\affiliation{Cavendish Laboratory, University of Cambridge, J.J. Thomson Ave, Cambridge CB3 0HE, United Kingdom}


\date{\today}

\begin{abstract}
Over the last two decades organic spintronics has developed into a striving field with exciting reports of long spin diffusion lengths and spin relaxation times in organic semiconductors (OSCs). Easily processed and inexpensive, OSCs are considered a potential alternative to inorganic materials for use in spintronic applications. Spin currents have been detected in a wide range of materials, however, there is still uncertainty over the origin of the signals. Recently, we explored spin transport through an organic semiconductor with lateral spin injection and detection architectures, where the injected spin current is detected non-locally via spin-to-charge conversion in an inorganic detector. In this work we show that the widely-used control experiments like linear power dependence and inversion of the signal with the magnetic field are not sufficient evidence of spin transport and can lead to an incorrect interpretation of the signal. Here, we use in-plane angular dependent measurements to separate pure spin signal from parasitic effects arising from spin rectification (SREs). Apart from well established anisotropic magnetoresistance (AMR) and anomalous Hall effect (AHE), we observed a novel effect which we call \emph{spurious} inverse spin Hall effect (ISHE). It strongly resembles ISHE behaviour, but arises in the ferromagnet rather than the detector meaning this additional effect has to be considered in future work.

\end{abstract}


\maketitle

\section{Introduction}

Organic semiconductors are a class of van der Waals bonded soft materials predominantly composed of light chemical elements such as carbon, hydrogen and sulphur. Owing to their composition, the spin orbit coupling within these materials is smaller compared to crystalline inorganic semiconductors, such as  silicon or gallium arsenide. The reduced spin-orbit coupling in organic semiconductors in turn manifests itself through long carrier spin lifetimes, typically on the order of several microseconds \cite{schott2017tuning},\cite{schott2019polaron}. Such durations are long enough for the coherent manipulation of quantum spin information to be carried out and opens up new uses for organic semiconductors in spintronics. One drawback of the intrinsic nature of van der Waals bonding and the lack of consistent long-range order in organic semiconductors, however, is that their charge carrier mobilities tend to be comparatively low and typically on the order of $\mathrm{1\ cm^{\mathrm{2}}V^{\mathrm{-1}}s^{\mathrm{-1}}}$. The immediate consequence of such low mobilities for spintronics is that the spin diffusion lengths within devices are expected to be small, as the spin diffusion length is computed using the relation $L_{\mathrm{S}}=\sqrt{D\tau_{\mathrm{S}}}$. Here $L_{\mathrm{S}}$ is the spin diffusion length, $\tau_{\mathrm{S}}$  is the carrier spin lifetime and $D$ is the diffusion constant linked to the carrier mobility $\mu$ with Einstein’s relationship $D=\mu k_{\mathrm{B}}T/q$. It thus comes as no surprise that the reported values of the spin diffusion lengths in several undoped organic semiconductors (e.g. $\mathrm{Alq_{\mathrm{3}}}$, P3HT, PPV and TIPS-Pentacene) were typically measured to be under 100 nm \cite{xiong2004giant},\cite{nguyen2010isotope},\cite{nguyen2012spin},\cite{mooser2012spin},\cite{li2015excellent},\cite{zhang2013observation},\cite{majumdar2012on},\cite{morley2008room}. It also becomes evident that the only way to boost the magnitude of the spin diffusion length beyond 100 nm in organic semiconductors is to amplify the diffusion constant $D$, for which purpose a new mechanism based on spin exchange transport was proposed \cite{yu2014suppression},\cite{yu2011spin},\cite{yu2014impurity}.

The sub-100 nm spin diffusion lengths mentioned above were measured using devices known as vertical organic spin-valves. Vertical organic spin valves sandwich an organic semiconductor layer between two ferromagnetic thin film electrodes and show a switching behaviour in the electrically measured magnetoresistance of the trilayer stack in response to an applied magnetic field. Organic spin valves are easy to fabricate and their magnetoresistance measurements are relatively straightforward to interpret. However, the following three recently identified issues call into question the unambiguity of organic spin valve-based measurement techniques. First, it has been suggested that an alternative mechanism for the magnetoresistance switching signals in the organic trilayer stack stems from biaxial anisotropy and Tunnelling Anisotropic Magneto Resistance (TAMR) of the ferromagnetic electrode itself \cite{grunewald2011tunneling}. Second, there has never been an unambiguous demonstration of the Hanle effect within organic spin valves \cite{grunewald2013vertical}. Third, a lack of scaling in the resistance-area products of some organic spin valve junctions casts light on the role pinholes play in a measured magnetoresistance signal \cite{gockeritz2016resistive}. Pinholes are understood to develop within the organic spin valve during device fabrication, when the top ferromagnetic electrode is deposited onto a soft organic sandwich layer.

\begin{figure}[t]
\includegraphics[width = \linewidth]{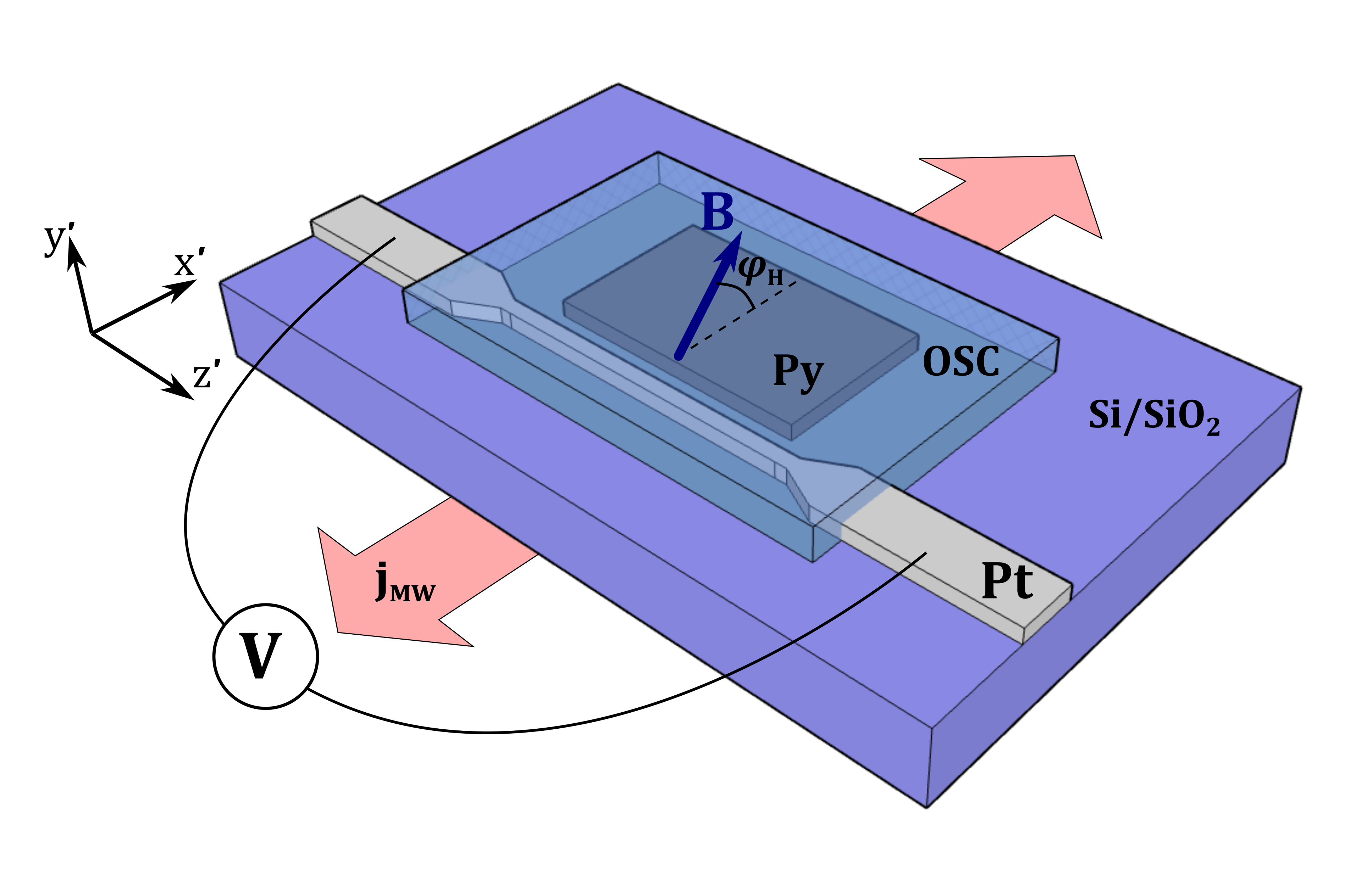}
\caption{\label{fig:sample_3d}A schematic diagram of the LSP sample architecture and the voltage measurement.}
\end{figure}

The above challenges in unambiguously performing spin transport measurements within organic semiconductor spin valves call for improved techniques. In the recent past, we attempted to overcome some of the aforementioned challenges by using a new all-electrical device-based method known as lateral spin pumping (LSP) \cite{wang2019long}. A schematic diagram of the lateral spin pumping device and the voltage measurement is shown in fig.~\ref{fig:sample_3d}. For spin current injection into the organic semiconductor a non-equilibrium spin current is generated by spin pumping using microwave excitation of ferromagnetic resonance within a thin film magnet in the device. The injected spin current is then assumed to propagate laterally through the organic semiconductor in contact with the thin film magnet. When the spin current arrives at a heavy metal stripe (platinum), which is positioned in proximity to the ferromagnetic spin injector but is only in direct contact with the organic semiconductor, it is converted into an electromotive force by the inverse spin Hall effect (ISHE). A measurable electromotive force is accepted evidence of spin transport within such a device. The device fabrication scheme incorporates the organic semiconductor in a final step thus ensuring no damage to the soft spin transport layer. The technique also overcame several associated intrinsic measurement difficulties presented by nonlocal spin valve devices that use organic semiconductors \cite{wang2019long}.

Lateral spin pumping is an extension of bilayer spin pumping that was first postulated in the late 1970s within non-magnet/ferromagnet stacks \cite{silsbee1979coupling}, where a spin current is generated in the non-magnet through the dissipation of angular momentum within a ferromagnet that it sits in contact with \cite{tserkovnyak2002enhanced}. However, it wasn’t until the mid 2000s that the first experimental attempt to validate the theory was made \cite{saitoh2006conversion}. This demonstration made use of a permalloy/platinum bilayer where the spin current was excited within permalloy and detected within platinum using the inverse spin Hall effect. The technique inspired similar measurements on permalloy/non-magnet bilayers and was used to estimate the spin Hall angle of the non-magnetic spin detector \cite{sinova2015spin}. In recent years however, the use of conductive ferromagnets such as permalloy within spin pumping devices has been discouraged, primarily due to a newly understood effect known as spin rectification. In spin rectification, a mixing signal between an rf generated current and the magnetoresistance in the conductive ferromagnet can yield a dc voltage component which can contribute to the signal and has similar symmetries as the signal generated by the inverse spin Hall effect \cite{iguchi2017measurement}.

In the past, lateral spin pumping using permalloy spin injectors was used to probe spin transport in a variety of materials including p-Si, \cite{shikoh2013spin} n-GaAs, \cite{yamamoto2015spin} graphene, \cite{tang2013dynamically} n-Ge, \cite{dushenko2015experimental} Cu, \cite{yamamoto2014characterization} and in a two-dimensional electron gas \cite{ohshima2017strong}. These studies were a basis for our own earlier work on lateral spin pumping into an organic semiconductor \cite{wang2019long}. The aforementioned studies, however, did not investigate the potential contributions to the measured signal from spin rectification (SRE) which can accompany the use of conductive ferromagnets and arise from spin rectification \cite{bai2013distinguishing},\cite{bai2013universal},\cite{harder2011analysis},\cite{soh2014an}.

In this work, we extend our earlier study and demonstrate how contributions from spin rectification can be assessed within an organic lateral spin pumping device that uses conductive ferromagnetic injectors. We demonstrate the need for additional angular dependent measurements of the lateral spin pumping signal and identify various components to the measured ‘spin signal’. In addition to angular dependent measurements on lateral spin pumping devices, we use various on-chip scaling approaches in both inorganic bilayers as well as devices which include an organic semiconductor, to systematically study how the components of spin rectification may be minimised in future experiments. Our study will be of interest to both the organic spintronics as well as the inorganic spintronics communities, as it highlights the need for greater scrutiny in the interpretation of such lateral spin pumping measurements. 

\section{Theory}
\subsection{Ferromagnetic resonance}

The dynamic behaviour of a ferromagnetic thin film can be modelled by the phenomenological Landau-Lifshitz-Gilbert equation \cite{gilbert2004a} which describes the precession of the magnetisation $\bm{M}$ under the influence of a magnetic field $\vec{H}$
\begin{equation}
\frac{d\bm{M}}{dt} = - \gamma \bm{M} \times \vec{H} + \frac{\alpha}{M_{\mathrm{s}}} \bm{M} \times \frac{d\bm{M}}{dt},
\label{eq:LLG}
\end{equation}
where $\gamma$, $\alpha$ and $M_{\mathrm{s}}$ are the gyromagnetic ratio, Gilbert damping parameter and saturation magnetisation, respectively. The magnetic field $\bm{H}$ includes contributions from the external magnetic field as well as the demagnetisation fields.

\begin{figure}[t]
\includegraphics[width = \linewidth]{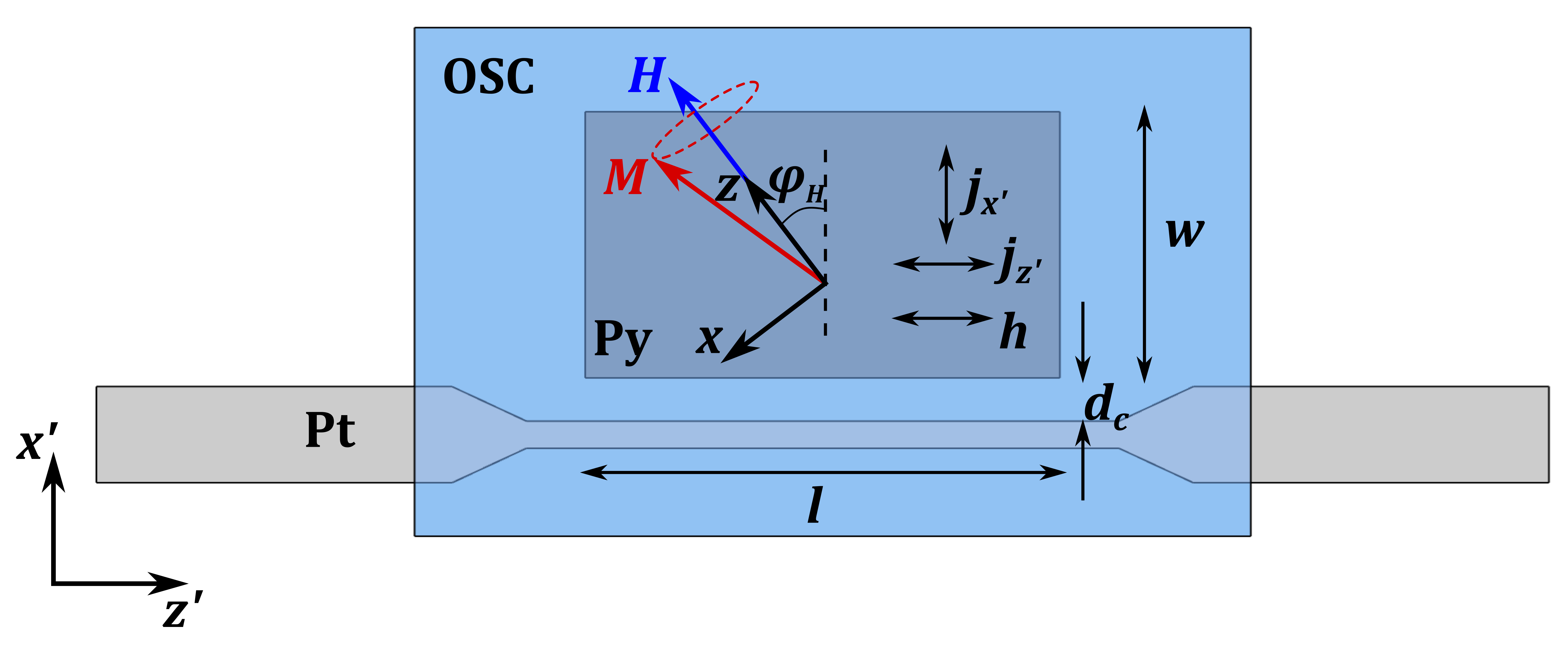}
\caption{\label{fig:sample_2d}A two-dimensional, top-view diagram of the LSP sample showing directions of the external field $\bm{H}$, magnetisation $\bm{M}$ as well as induced microwave currents $\bm{j_{\mathrm{x'}}}$, $\bm{j_{\mathrm{z'}}}$ and magnetic field $\bm{h}$.}
\end{figure}

To include the effect of the microwave magnetic field $he^{-i \omega t}$ oscillating at a frequency $\omega$ and analyse the in-plane angular dependence, we follow a similar procedure as in \cite{soh2014an} and define a coordinate system $\left(x,y,z\right)$ which rotates together with the magnetisation $\bm{M}$ aligned with the $z$ axis, as shown in fig.~\ref{fig:sample_2d}. The coordinates of the rotating frame $(x, y, z)$ and the laboratory frame $(x', y', z')$ are related by a rotation matrix $R_{\mathrm{y}}(\pi /2 + \varphi_{\mathrm{H}})$ about the $y$ axis, where $\varphi_{\mathrm{H}}$ is the in-plane angle between the magnetic field $\bm{H}$ and the direction of the microwave waveguide, which is the $x'$ axis. Solving eq.~\ref{eq:LLG} for the ac magnetisation $me^{\mathrm{-i\omega t}}$ yields the dynamic susceptibility tensor $\chi$ which relates the oscillating magnetisation $\bm{m}$ and magnetic field $\bm{h}$ in the following way:
\begin{equation}
\bm{m} = \bm{R_{\mathrm{y}}}(\pi /2 + \varphi_{\mathrm{H}})\; \bm{\chi}\; \bm{R_{\mathrm{y}}}(- \pi /2 - \varphi_{\mathrm{H}}),
\label{eq:m}
\end{equation}
\begin{equation}
\bm{\chi} = \begin{pmatrix}
A_{\mathrm{xx}} & iA_{\mathrm{xy}} & 0 \\
iA_{\mathrm{xy}} & A_{\mathrm{yy}} & 0 \\
0 & 0 & 0
\end{pmatrix},
\label{eq:chi}
\end{equation}
\begin{subequations}
\label{eq:As_whole}
\begin{equation}
A_{\mathrm{xy}} = \frac{-4\pi M_{\mathrm{s}}}{\alpha(2H_{\mathrm{0}} + 4\pi M_{\mathrm{s}})}, 
\label{eq:Axy}
\end{equation}
\begin{equation}
A_{\mathrm{xx}} = - \frac{A_{\mathrm{xy}}\gamma(H_{\mathrm{0}} + 4\pi M_{\mathrm{s}})}{\omega},
\label{eq:Axx}
\end{equation}
\begin{equation}
A_{\mathrm{yy}} = - \frac{A_{\mathrm{xy}}\gamma H_{\mathrm{0}}}{\omega}.
\label{eq:Ayy}
\end{equation}
\end{subequations}

The behaviour of ferromagnetic resonance (FMR) is captured by L and D which define the symmetric and antisymmetric lineshape components:
\begin{subequations}
\label{eq:lineshapes}
\begin{equation}
\mathrm{L} = \frac{\Delta H^{\mathrm{2}}}{4(H_{\mathrm{0}} - H_{\mathrm{r}})^{\mathrm{2}} + \Delta H^{\mathrm{2}}},
\label{eq:L}
\end{equation}
\begin{equation}
\mathrm{D} = \frac{2\Delta H (H_{\mathrm{0}} - H_{\mathrm{r}})}{4(H_{\mathrm{0}} - H_{\mathrm{r}})^{\mathrm{2}} + \Delta H^{\mathrm{2}}}.
\label{eq:D}
\end{equation}
\end{subequations}
In the above equations $H_{\mathrm{r}}$ and $\Delta H$ are the resonant field value and the resonance linewidth (full-width at half-maximum).

\subsection{Spin pumping and inverse spin-Hall effect}

Let’s consider a Py/Pt bilayer under the FMR condition. Magnetisation dynamics causes spin angular momentum transfer from Py to Pt in the form of a spin current. The spin current is converted into a charge current via ISHE
\begin{equation}
\bm{j_{\mathrm{c}}} = \frac{2e\theta_{\mathrm{SHE}}}{\hbar} \hat{y} \times \bm{j_{\mathrm{s}}},
\label{eq:ishe}
\end{equation}
where $\bm{j_{\mathrm{s}}}$ is the spin current averaged over the Pt thickness and takes the form \cite{ando2011inverse}
\begin{equation}
\bm{j_{\mathrm{s}}} = \frac{\hbar \lambda_{\mathrm{Pt}} g_{\mathrm{eff}}^{\mathrm{\uparrow \downarrow}}}{4\pi M_{\mathrm{0}}^{\mathrm{2}} d_{\mathrm{Pt}}} \mathrm{tanh}\left(\frac{d_{\mathrm{Pt}}}{2\lambda_{\mathrm{Pt}}}\right)Re\bigg\{\bm{M} \times \frac{d\bm{M}}{dt}\bigg\}.
\label{eq:js}
\end{equation}
Here, $M_{\mathrm{0}}$ is the magnitude of equilibrium magnetisation, $\lambda_{\mathrm{Pt}}$ is the spin diffusion length in Pt, $d_{\mathrm{Pt}}$ is the thickness of the Pt film and $g_{\mathrm{eff}}^{\mathrm{\uparrow\downarrow}}$ is the real part of the effective spin mixing conductance. By substituting the solution to eq.~\ref{eq:m}, time averaged and scaled by the resistance of the bilayer, we can obtain the final form of the DC component of the ISHE voltage
\begin{equation}
V_{\mathrm{ISHE}}^{\mathrm{DC}} = -f_{\mathrm{Pt}} K A_{\mathrm{xy}}A_{\mathrm{xx}}h^{\mathrm{2}}\omega \mathrm{cos}^{\mathrm{3}}\varphi_{\mathrm{H}} \cdot \mathrm{L},
\label{eq:Vishe}
\end{equation}
where $K=we\theta_{\mathrm{SHE}}g_{\mathrm{eff}}^{\mathrm{\uparrow\downarrow}} \lambda_{\mathrm{Pt}} \mathrm{tanh}(d_{\mathrm{Pt}}/2\lambda_{\mathrm{Pt}})/64\pi^{\mathrm{3}}M_{\mathrm{s}}^{\mathrm{2}}$ and $f_{\mathrm{Pt}}=1/(d_{\mathrm{Pt}}\sigma_{\mathrm{Pt}} + d_{\mathrm{Py}}\sigma_{\mathrm{Py}})$, with $d_{\mathrm{Pt}}$, $d_{\mathrm{Py}}$ and $\sigma_{\mathrm{Pt}}$, $\sigma_{\mathrm{Py}}$ being the thickness and conductivity values for Pt and Py layers, respectively.

\subsection{Spin rectification effects}

To analyse the angular dependence of the SRE, we start by considering the generalised Ohm’s law \cite{juretschke1960electromagnetic} which couples together the current density $\bm{J}$, magnetisation $\bm{M}$, electric field $\bm{E}$ and magnetic field $\bm{H}$ inside the ferromagnet
\begin{equation}
\bm{J} = \sigma\bm{E} - \frac{\sigma \Delta \rho}{M^{\mathrm{2}}}(\bm{J}\cdot\bm{M})\bm{M} + \sigma R_{\mathrm{H}} (\bm{J}\times\bm{H}) + \sigma R_{\mathrm{AHE}}(\bm{J}\times\bm{M}).
\label{eq:ohm}
\end{equation}
The terms in this equation represent the usual Ohmic contribution, anisotropic magnetoresistance (AMR), standard Hall effect and anomalous Hall effect (AHE) contributions, respectively. $\sigma$, $\Delta\rho$, $R_{\mathrm{H}}$ and $R_{\mathrm{AHE}}$ are the conductivity, change in resistivity due to AMR, ordinary and anomalous Hall coefficients. Time-averaging the above expression yields a DC current
\begin{eqnarray}
\bm{J}_{\mathrm{SRE}} =&&\ -\ \frac{\sigma \Delta \rho}{M^{\mathrm{2}}}\left(\langle\bm{j}\times\bm{m}\rangle\times \bm{M} + \langle\bm{j}\cdot\bm{m}\rangle\bm{M}\right)\nonumber\\
&&\ +\ \sigma R_{\mathrm{H}} \langle\bm{j}\times\bm{h}\rangle + \sigma R_{\mathrm{AHE}}\langle\bm{j}\times\bm{m}\rangle,
\label{eq:jsre}
\end{eqnarray}
where $\bm{j}$, $\bm{h}$ and $\bm{m}$ are the ac components of the microwave-induced current, magnetic field and magnetisation. These three quantities oscillate at the same frequency $\omega$ but with different phases. Specifically, $\bm{j}$ and $\bm{h}$ oscillate with a relative phase $\Phi$ which is usually referred to as the electromagnetic phase difference. The usual Hall contribution being independent of magnetisation results in a constant offset, therefore will be ignored in further analysis.

In our geometry the microwave current $\bm{j}$ flows mostly in the $x'$ direction. However, due to the complexity of transmission line-to-sample coupling and possible sample misalignment on the CPW, we allow for the current to have two in-plane components $\bm{j}_{\mathrm{x'}}$ and $\bm{j}_{\mathrm{z'}}$. Using the solution to eq.~\ref{eq:m} we obtain the final expression for the SRE voltage signal along the measurement direction $z'$:
\begin{subequations}
\label{eq:sre}
\begin{equation}
V_{\mathrm{SRE}} = A_{\mathrm{L}} \cdot \mathrm{L} + A_{\mathrm{D}} \cdot \mathrm{D},
\label{eq:Vsre}
\end{equation}
\begin{eqnarray}
A_{\mathrm{L}} =&& sin \Phi \cdot (V_{\mathrm{AMR}}^{\mathrm{x}} \mathrm{cos}\varphi_{\mathrm{H}} \mathrm{cos}2\varphi_{\mathrm{H}} - V_{\mathrm{AMR}}^{\mathrm{z}} \mathrm{cos}\varphi_{\mathrm{H}} \mathrm{sin}2\varphi_{\mathrm{H}})\nonumber\\
&&-\mathrm{cos}\Phi \cdot V_{\mathrm{AHE}} \mathrm{cos}\varphi_{\mathrm{H}},
\label{eq:Al}
\end{eqnarray}
\begin{eqnarray}
A_{\mathrm{D}} =&& -cos \Phi \cdot (V_{\mathrm{AMR}}^{\mathrm{x}} \mathrm{cos}\varphi_{\mathrm{H}} \mathrm{cos}2\varphi_{\mathrm{H}} - V_{\mathrm{AMR}}^{\mathrm{z}} \mathrm{cos}\varphi_{\mathrm{H}} \mathrm{sin}2\varphi_{\mathrm{H}})\nonumber\\
&&- \mathrm{sin}\Phi \cdot V_{\mathrm{AHE}} \mathrm{cos}\varphi_{\mathrm{H}},
\label{eq:Ad}
\end{eqnarray}
\end{subequations}
where L and D are the symmetric and antisymmetric lineshapes defined in eqs.~\ref{eq:lineshapes}.

The voltage contributions due to AMR from microwave currents $\bm{j}_{\mathrm{x'}}$ and $\bm{j}_{\mathrm{z'}}$, and due to AHE are:
\begin{subequations}
\label{eq:sre-voltages}
\begin{equation}
V_{\mathrm{AMR}}^{\mathrm{x}} = f_{\mathrm{Py}}A_{\mathrm{xx}}hj_{\mathrm{x'}}\Delta\rho w/8\pi M_{\mathrm{s}},
\end{equation}
\begin{equation}
V_{\mathrm{AMR}}^{\mathrm{z}} = f_{\mathrm{Py}}A_{\mathrm{xx}}hj_{\mathrm{z'}}\Delta\rho w/8\pi M_{\mathrm{s}},
\end{equation}
\begin{equation}
V_{\mathrm{AMR}}^{\mathrm{x}} = f_{\mathrm{Py}}A_{\mathrm{xy}}R_{\mathrm{AHE}}j_{\mathrm{x'}}hw/2.
\end{equation}
\end{subequations}
The factor $f_{\mathrm{Py}}=d_{\mathrm{Py}}\sigma_{\mathrm{Py}}/\left(d_{\mathrm{Py}}\sigma_{\mathrm{Py}}+d_{\mathrm{Pt}}\sigma_{\mathrm{Pt}}\right)$ incorporates shorting of the SRE voltage contributions by the Pt layer.

One can see that the lineshape of SREs is strongly dependent on the electromagnetic phase $\Phi$. It is a material and frequency dependent parameter related to the losses in the system – an electromagnetic wave propagating through an absorptive medium has a complex wave vector, whose imaginary contribution creates a phase shift between electric and magnetic fields. It is difficult to compute $\Phi$ in a system such as ours with a complex coupling between the CPW and a metallic sample, and therefore it is treated as one of the fitting parameters. Previous work \cite{harder2011analysis} showed that it can change significantly within a small range of microwave frequencies and between different samples.

\begin{figure}[b]
\includegraphics[width = 0.9\linewidth]{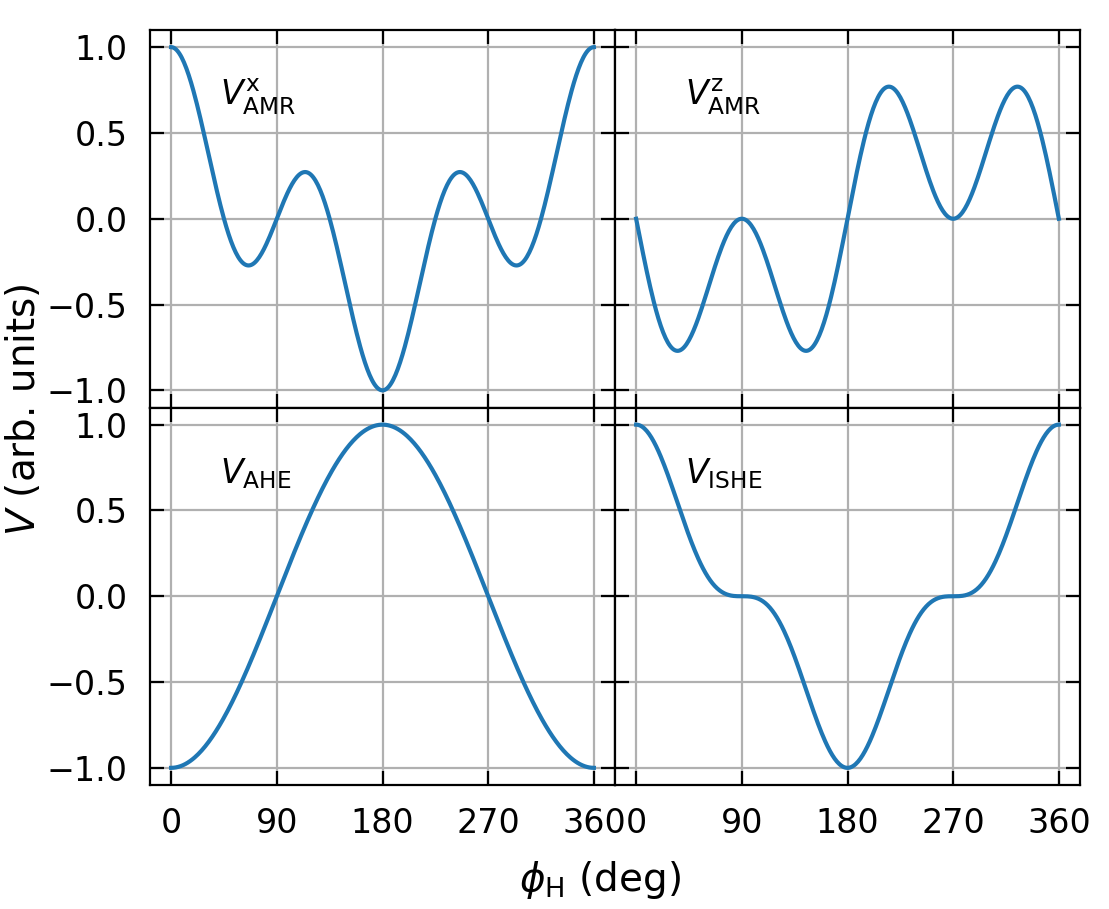}
\caption{\label{fig:ang_dep_shapes}Modelled in-plane angular dependence of the ISHE signal and the three SRE contributions: $\mathrm{AMR^{\mathrm{x}}}$, $\mathrm{AMR^{\mathrm{z}}}$ and AHE.}
\end{figure}

In general, each SRE component can create both a symmetric and an antisymmetric voltage signal, which makes it impossible to distinguish ISHE from SRE based solely on the lineshape symmetry. However, they can be separated based on the dependence on the in-plane angle $\varphi_{\mathrm{H}}$. Figure \ref{fig:ang_dep_shapes} shows the expected angular dependence profiles for each voltage component. This forms the basis of our analysis in this paper - after extracting $A_{\mathrm{L}}$ and $A_{\mathrm{D}}$ values from a magnetic field sweep at each angle, we then fit eqs.~\ref{eq:Vishe} and \ref{eq:sre} to extract the four contributions: $V_{\mathrm{AMR}}^{\mathrm{x}}$, $V_{\mathrm{AMR}}^{\mathrm{z}}$, $V_{\mathrm{AHE}}$ and $V_{\mathrm{ISHE}}$. This analysis applies directly to a bilayer system. In the LSP geometry the prefactors in eq.~\ref{eq:sre-voltages} containing the dependence on the geometric dimensions and the conductivity of the different materials would need to be adjusted, but the angular dependence analysis remains valid.

\begin{figure}[t]
\includegraphics[width = 0.6\linewidth]{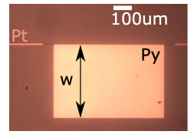}
\caption{\label{fig:micrograph}A micrograph of the device showing the central region before deposition of the OSC.}
\end{figure}

\section{Methods}

\subsection{Device fabrication}
All of the devices were fabricated on 3 mm x 5 mm Si/Si$\mathrm{O_{\mathrm{2}}}$ (300 nm) substrates. The lateral spin pumping devices were produced with the multistep ebeam lithography method used in reference \cite{wang2019long}. Firstly, a 10 nm thick, 1 $\mathrm{\mu m}$ wide Pt stripe was defined with electron beam lithography and subsequently deposited using magnetron sputtering at 1 x $\mathrm{10^{\mathrm{-6}}}$ mbar base pressure and 5 x $\mathrm{10^{\mathrm{-4}}}$ partial Ar gas pressure at 4.5 \AA/s rate. Secondly, a 600 $\mathrm{\mu m}$ x 400 $\mathrm{\mu m}$ x 25 nm Py injector was evaporated using an electron beam at 2 \AA/s under 2 x $\mathrm{10^{\mathrm{-7}}}$ mbar pressure. It was aligned parallel to the Pt stripe at a distance of 200 nm away from it by a second electron beam lithography step. Figure \ref{fig:micrograph} shows a microscope image of the active area of the LSP device at this stage in the fabrication process. Thirdly, the conjugated polymer poly[2,5-bis(3-tetradecylthiophen-2-yl)thieno[3,2-b]thiophene], also called PBTTT, was spin coated on top from a 10 g/l DCB solution at 5000 rpm in nitrogen atmosphere. The film was annealed at $\mathrm{180^{\mathrm{\circ}}C}$ for 20 min and slowly cooled back to room temperature to form the terraced phase. Finally, the devices were placed in 1 g/l F4TCNQ:ACN solution to dope the OSC. They were then annealed at $\mathrm{80^{\mathrm{\circ}}C}$ for 20 min to allow the dopant to diffuse into the film. The estimated conductivity achieved for these films was around 100 $\mathrm{Scm^{\mathrm{-1}}}$. The doped OSC film was patterned into a box covering the Py injector, the channel, and the Pt stripe by scratching the excess with a cocktail stick. The stripe was attached to wider, 10 $\mathrm{\mu m}$ Pt pads which were used to contact the device with silver dag.

To fabricate the bilayer devices we deposited the bottom contact (Au or Pt) through a 2 mm x 4 mm shadow mask. Au was thermally evaporated at 1 x $\mathrm{10^{\mathrm{-6}}}$ mbar base pressure while Pt was sputtered under the same conditions as the LSP devices above. Py was subsequently deposited on top of the contact through a 2 mm x 0.5 mm shadow mask using e-beam evaporation under identical conditions to the LSP devices.

\subsection{Measurement setup}
All of the device voltages were measured using a coplanar waveguide (CPW) based spin pumping setup with an in-plane magnetic field rotation. A magnetic field was applied in the plane of the device at different angles $\varphi_{\mathrm{H}}$ to magnetise the FM film. An in-plane 4 GHz microwave field was applied using a CPW with a 700 $\mathrm{\mu m}$ wide central conductor. The devices were contacted with Ag paste and placed face down on the CPW to maximise power absorption. To accurately measure the ferromagnetic absorption, a lock-in measurement technique was employed with magnetic field modulation at 20 Hz and modulation amplitude approximately 0.1 mT. To this end, one side of the CPW was connected to an Stanford Research Systems SR860 DSP lock-in amplifier through a rectifying diode. The device voltage was measured using a KEITHLEY 2182A Nanovoltmeter. The measurements were performed at $\mathrm{5^{\mathrm{\circ}}}$ steps as the magnetic field was rotated in-plane from $\varphi_{\mathrm{H}}$=$\mathrm{180^{\mathrm{\circ}}}$ to $\mathrm{360^{\mathrm{\circ}}}$.

\subsection{Data collection and analysis}
At each magnetic field angle $\varphi_{\mathrm{H}}$ both the device voltage and microwave absorption were measured as a function of the magnetic field H to capture the FMR. Each voltage scan was then fitted numerically with a combination of lineshapes in eq.~\ref{eq:lineshapes} to obtain best fit values of $V_{\mathrm{L}}$ and $V_{\mathrm{D}}$, corresponding to the magnitudes of the symmetric and antisymmetric Lorentzian lineshapes. The absorption scans were fitted with derivatives of the two lineshapes to extract $P_{\mathrm{L}}$ and $P_{\mathrm{D}}$ amplitudes, which allowed us to quantify the absorbed power as $P_{\mathrm{L}}^{\mathrm{2}}$. Next, each voltage component was plotted as a function of the angle $\varphi_{\mathrm{H}}$ and a second numerical fitting step was performed where $V_{\mathrm{L}}$ and $V_{\mathrm{D}}$ angular dependencies were fitted simultaneously with a combination of eqs.~\ref{eq:Vishe} and \ref{eq:sre}. This allowed us to extract the best fit values for $\Phi$, $V_{\mathrm{AMR}}^{\mathrm{x}}$, $V_{\mathrm{AMR}}^{\mathrm{z}}$, $V_{\mathrm{AHE}}$ and $V_{\mathrm{ISHE}}$.

\section{Lateral spin pumping}

\begin{figure*}[t]
\begin{subfigure}{0.42\linewidth}
\caption{}
\includegraphics[width = \linewidth]{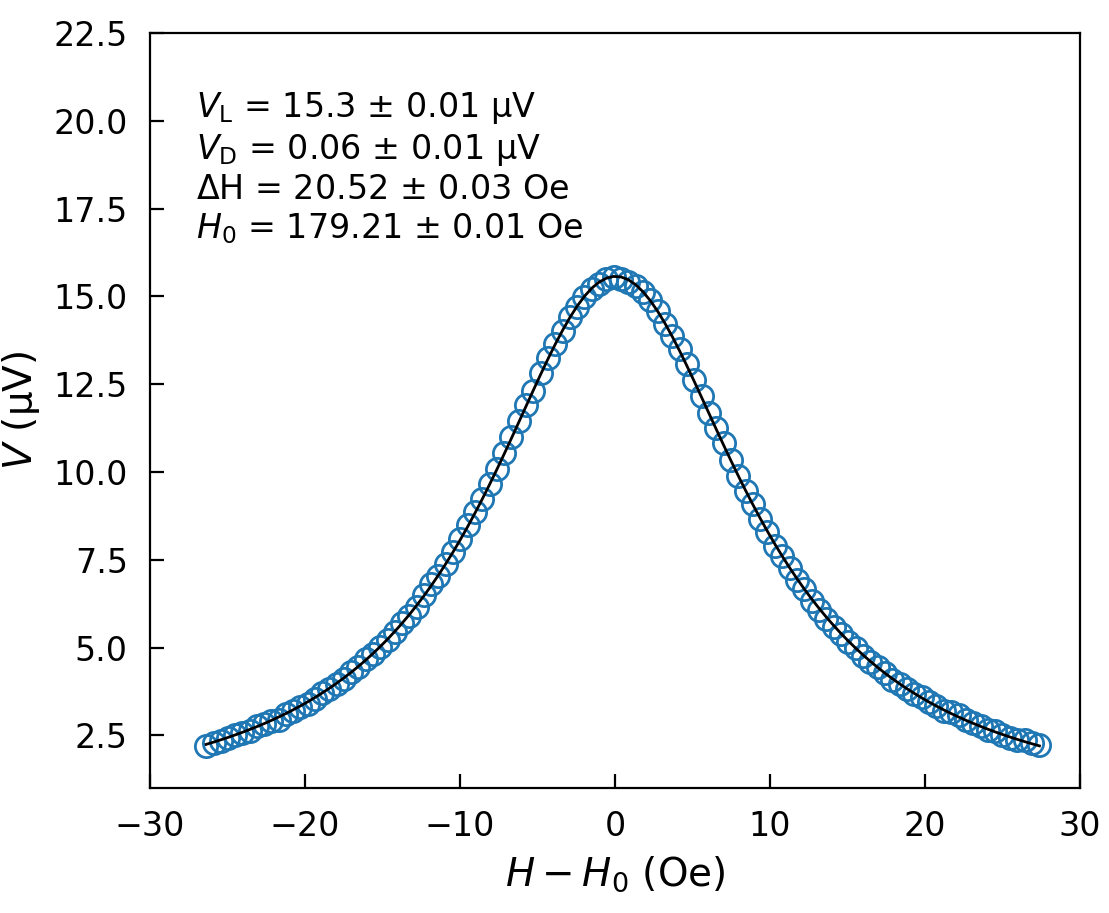}
\end{subfigure}
\begin{subfigure}{0.42\linewidth}
\caption{}
\includegraphics[width = \linewidth]{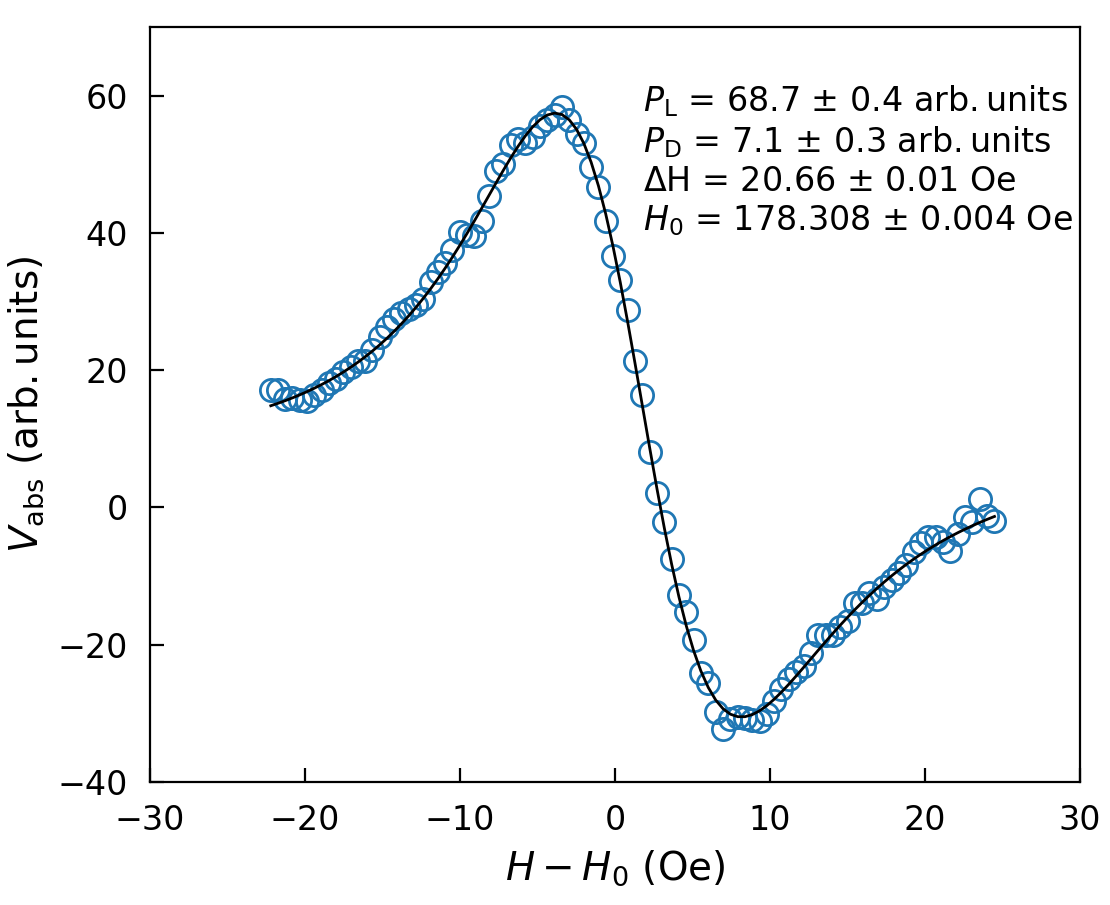}
\end{subfigure}
\begin{subfigure}{0.42\linewidth}
\caption{}
\includegraphics[width = \linewidth]{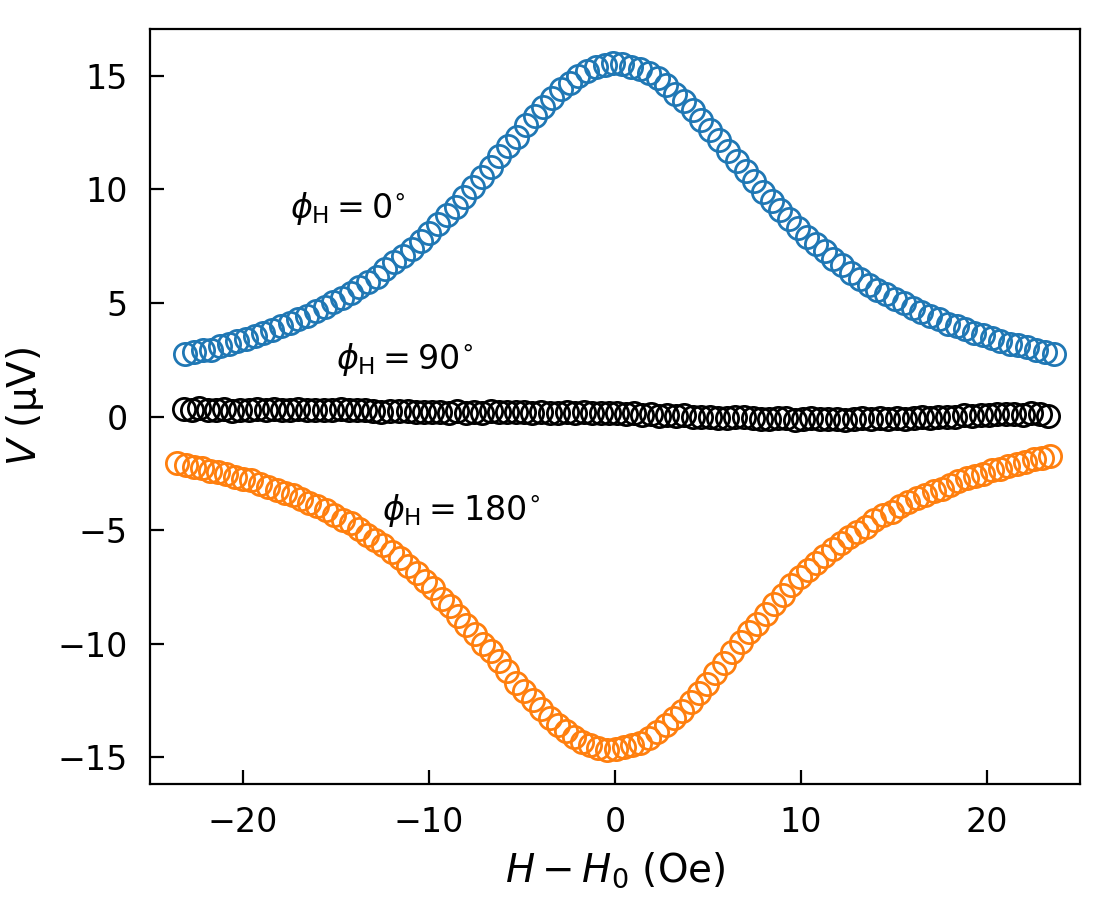}
\end{subfigure}
\begin{subfigure}{0.42\linewidth}
\caption{}
\includegraphics[width = \linewidth]{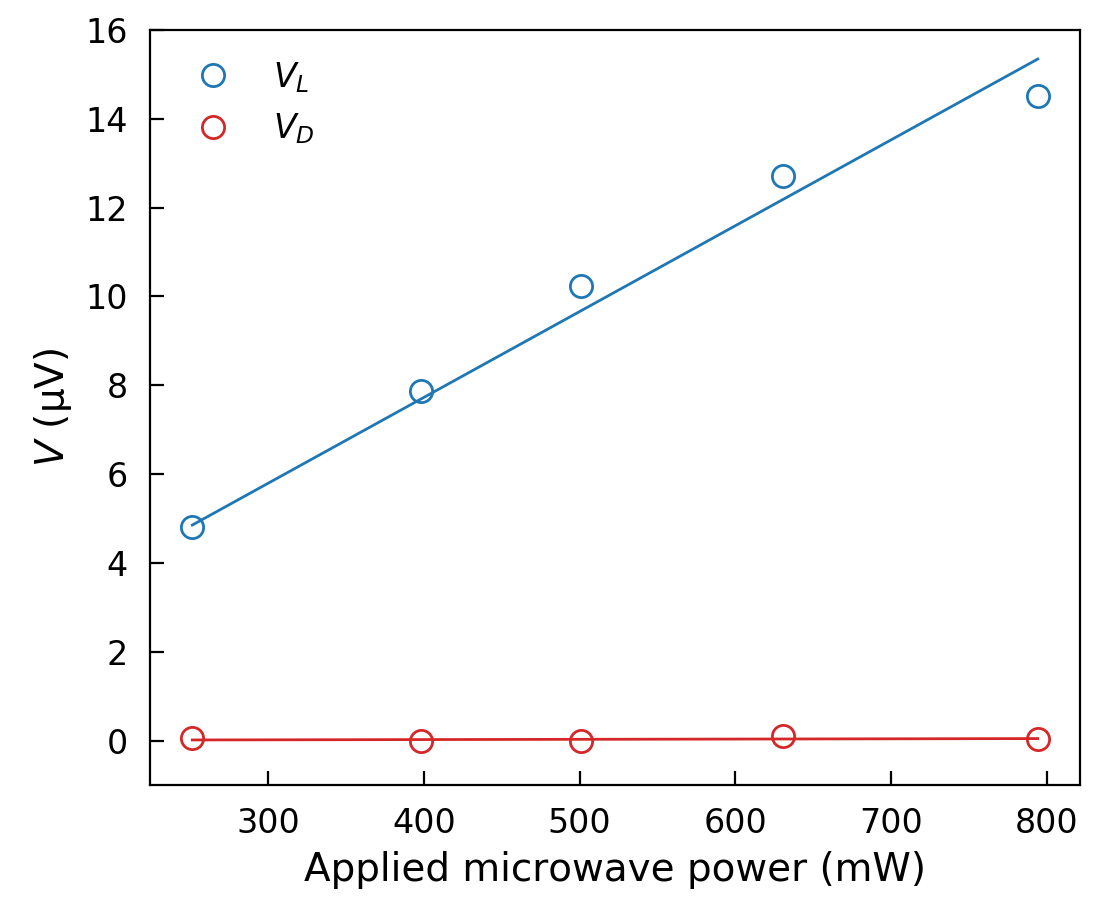}
\end{subfigure}
\caption{\label{fig:LSP1} LSP device with F4TCNQ-doped PBTTT and a 200 nm channel length between the Py injector and Pt detector. Voltage signal measured across the device (a) and the FMR absorption amplitude from the lock-in amplifier (b) measured at $\mathrm{\mathnormal{\varphi}_{H}=0^{\circ}}$ as a function of the applied magnetic field. (c) Voltage signal measured at three orientations of the magnetic field: $\mathrm{\mathnormal{\varphi}_{H}=0^{\circ},\ 90^{\circ}\ and\ 180^{\circ}}$. (d) Power dependence of the voltage signal at $\mathrm{\mathnormal{\varphi}_{H}=0^{\circ}}$.}
\end{figure*}

\begin{figure}[t]
\begin{subfigure}{0.9\linewidth}
\caption{}
\includegraphics[width = \linewidth]{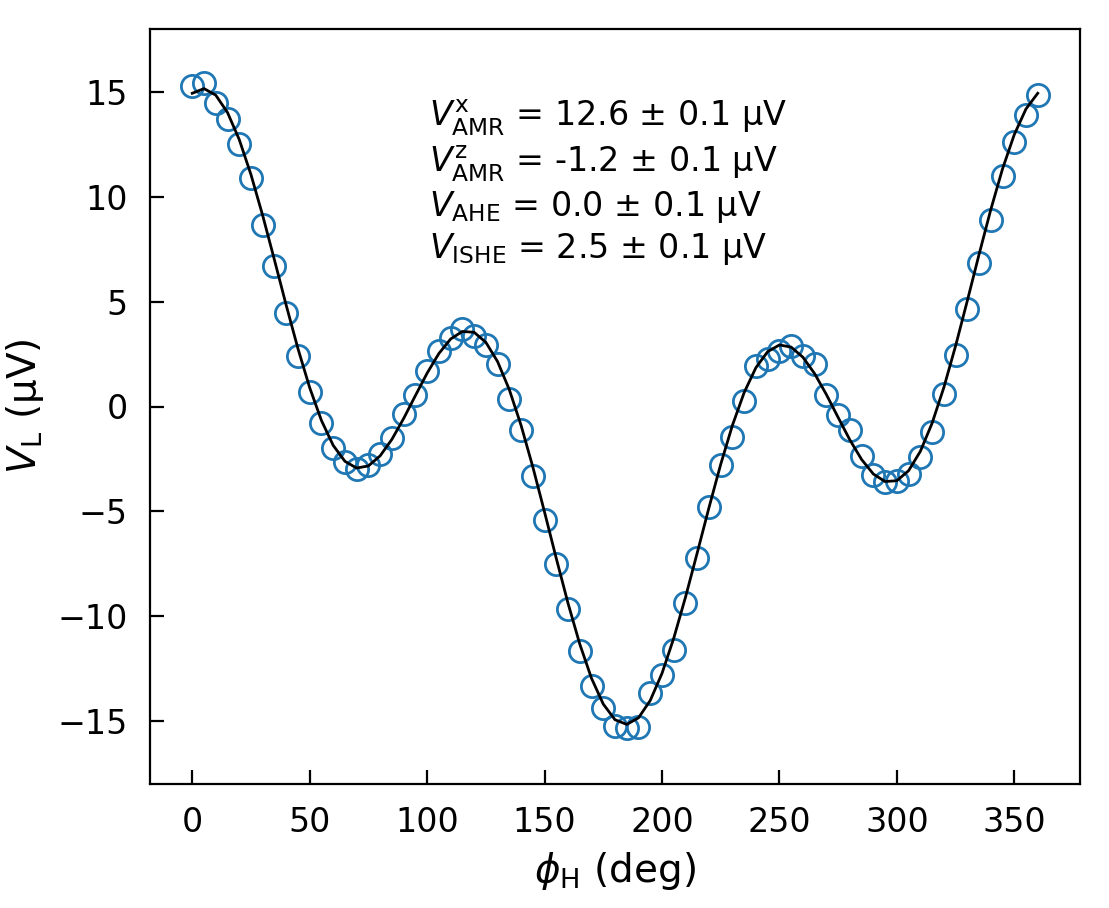}
\end{subfigure}
\begin{subfigure}{0.9\linewidth}
\caption{}
\includegraphics[width = \linewidth]{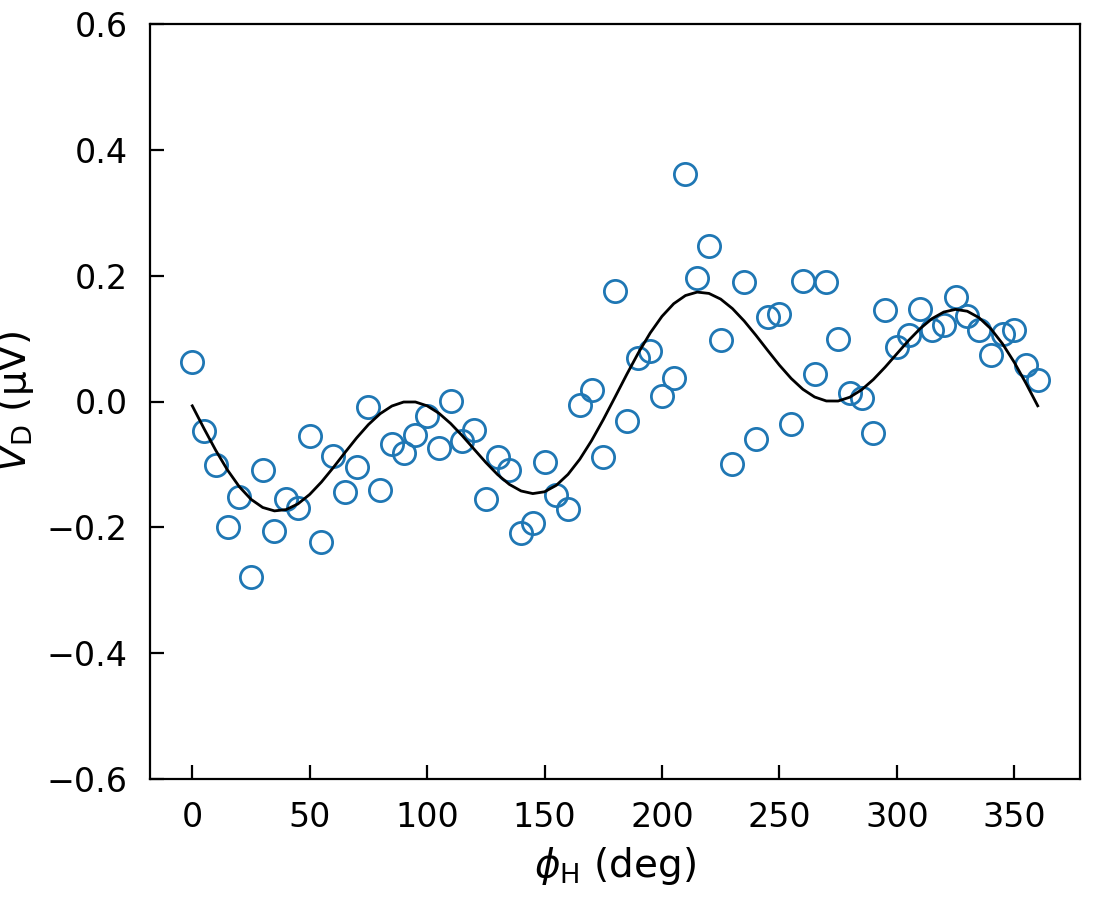}
\end{subfigure}
\caption{\label{fig:LSP2} LSP device with F4TCNQ-doped PBTTT and a 200 nm channel length between Py injector and Pt detector. In-plane angular dependence of (a) symmetric and (b) antisymmetric voltage components, respectively. Solid lines show numerical fits.}
\end{figure}

We produced a LSP device with F4TCNQ-doped PBTTT film and 200 nm channel length. In line with previous experiments on spin pumping, we measured the device voltage as a function of magnetic field at three critical in-plane angles: $\varphi_{\mathrm{H}}\mathrm{=0^{\mathrm{\circ}}}$, $\mathrm{{90}^{\mathrm{\circ}}}$ and $\mathrm{{180}^{\mathrm{\circ}}}$, and produced a microwave power dependence plot. Our results are shown in figs.~\ref{fig:LSP1} and \ref{fig:LSP2}. The voltage profile in fig.~\ref{fig:LSP1}(a) can be well fitted with a combination of a symmetric and an antisymmetric Lorentzian lineshapes, and its peak aligns with the position of the FMR absorption (fig.~\ref{fig:LSP1}(b)) at 179 Oe. In literature, the symmetric Lorentzian contribution has been attributed to the ISHE, while the antisymmetric Lorentzian contribution has simply been identified as coming from the AHE. Moreover, from fig.~\ref{fig:LSP1}(c) it can be seen that the voltage signal disappears at $\varphi_{\mathrm{H}}\mathrm{=0^{\mathrm{\circ}}}$ and has an opposite sign at $\varphi_{\mathrm{H}}\mathrm{={180}^{\mathrm{\circ}}}$ - this symmetry argument, while being in agreement with ISHE, has often been used to confirm the origin of the spin pumping voltage.

We argue, however, based on the angular dependence analysis of SREs that this simple symmetry argument is not sufficient to identify the origin of the voltage signal. Firstly, according to eq.~\ref{eq:sre}, rectified AHE as well as AMR can contribute to both symmetric and antisymmetric parts of the voltage signal. Only in certain special case can the symmetric part be free from spurious effects: either $\mathrm{\Phi=0^{\mathrm{\circ}}}$ and $\mathrm{\mathnormal{V}_{AHE}=0}$, or $\mathrm{\Phi={90}^{\circ}}$ and $\mathrm{\mathnormal{V}_{AMR}^{x}=\mathnormal{V}_{AMR}^{z}=0}$. Secondly, as shown in fig.~\ref{fig:ang_dep_shapes}, both AHE and $\mathrm{AMR^{\mathrm{x}}}$ reduce to zero at $\mathrm{\mathnormal{\varphi}_{H}={90}^{\circ}}$ and have an opposite sign at $\mathrm{\mathnormal{\varphi}_{H}={180}^{\circ}}$, therefore the results from fig.~\ref{fig:LSP1}(c) are consistent not only with ISHE but also with the expected AHE and $\mathrm{AMR^{x}}$ symmetries. Moreover, the linear power dependence of the voltage signal, fig.~\ref{fig:LSP1}(d), has been used in literature to further support the symmetry argument. The SREs, however, are also expected to have a linear power dependence. Both AHE and AMR can be expressed as a product of $\bm{m}$ with either $\bm{h}$ or $\bm{j}$, each of these three quantities has a square root dependence on the microwave power. Therefore, the end effect is linearly proportional to the absorbed power.

Nevertheless, the SRE and ISHE can be distinguished from one another by measuring a full in-plane angular dependence, since each of them has a distinct $\varphi_{\mathrm{H}}$ dependence as pictured in fig.~\ref{fig:ang_dep_shapes}. Therefore, we performed a full in-plane angular dependence measurement on the same device. Results and the corresponding numerical fits are plotted in fig.~\ref{fig:LSP2}. Various contributions to the spin pumping voltage were extracted with the largest one being $\mathrm{\mathnormal{V}_{AMR}^{x}=12.6\ \mu V}$ followed by $\mathrm{\mathnormal{V}_{ISHE}=2.5\ \mu V}$. This finding contradicts the earlier assumption that the SP voltage measured in an LSP architecture originates purely from the ISHE and signifies the importance of the in-plane angular dependence analysis. It also shows that the dominant spurious effect is AMR rather than AHE, which is consistent with similar recent experiments that make use of conductive ferromagnets as spin injectors \cite{bai2013distinguishing},\cite{bai2013universal},\cite{harder2011analysis},\cite{soh2014an}. 

Considering that the extracted $V_{\mathrm{ISHE}}$ component was a small contribution to the overall voltage signal, we performed further control experiments to confirm or rule out the existence of ISHE in spin pumping device architectures.

\subsection{Discontinuous Pt electrode experiment}

\begin{figure}[t]
\includegraphics[width = 0.9\linewidth]{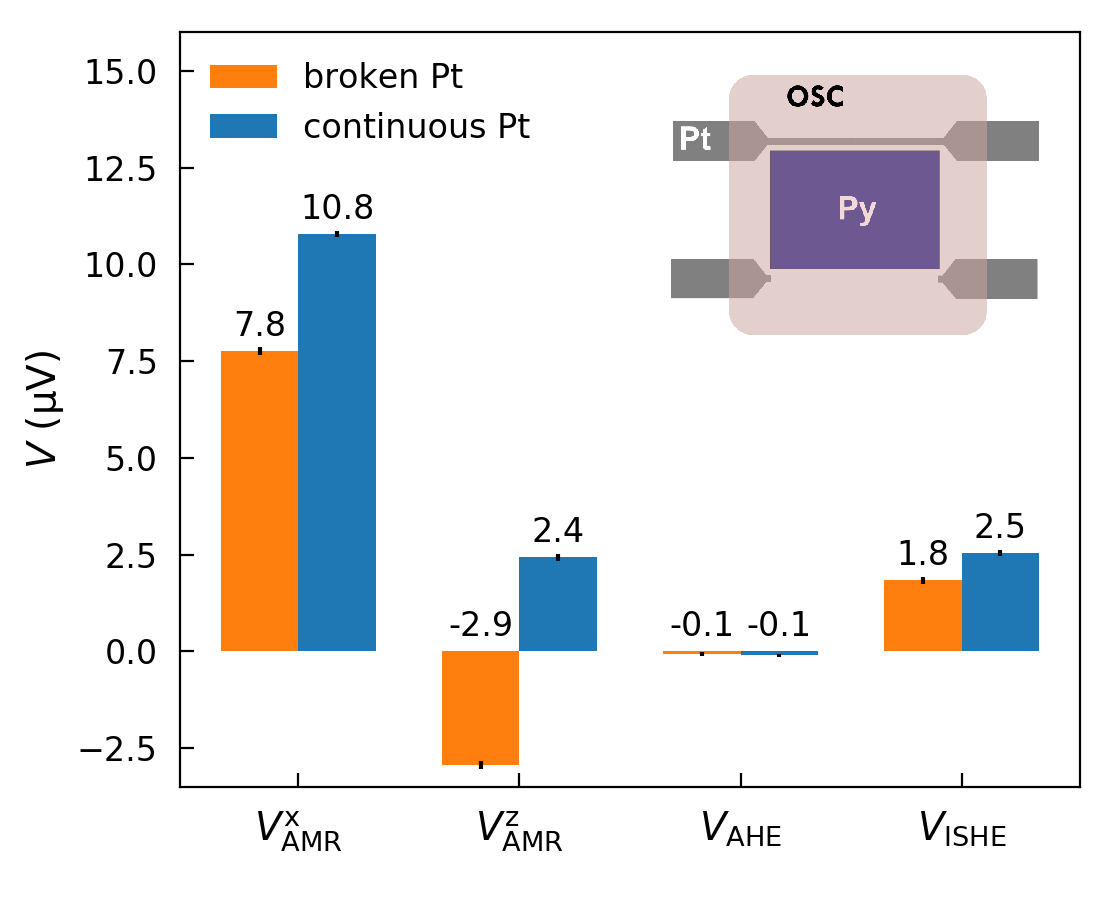}
\caption{\label{fig:broken_pt} Discontinuous platinum electrode experiment: comparison of the in-plane angular dependence voltage components for a LSP device with integrated continuous and discontinuous Pt electrodes on both sides of the Py injector (see inset).}
\end{figure}

The LSP device architecture was modified to include a second Pt electrode on the opposite side of the Py injector, as shown in fig.~\ref{fig:broken_pt} (inset). This additional Pt electrode was intentionally discontinuous and only extended 3 $\mathrm{\mu m}$ on either side next to the 600 $\mathrm{\mu m}$ long Py injector. With this architecture, in theory, the standard Pt electrode would convert the spin current into a charge current via the ISHE. It would also pick up any spurious voltage generated by the Py injector. Since the Pt stripe is missing on the other side, the discontinuous electrode, very little ISHE voltage is expected to be observed across this electrode, however any spurious voltage from Py would still be measurable.

Putting both Pt electrodes on the same device allowed us to control for device-to-device variations in Py/OSC qualities and hence spin injection/transport properties. Furthermore, it ensured that the contributions to the device resistance from the organic semiconductor and the Py injector were the same, making the resistance across both electrodes comparable.

The resulting voltage contributions measured across each electrode separately are summarised in fig.~\ref{fig:broken_pt}. Since the device had to be remounted between the two measurements, we normalised the voltages across the two electrodes by power absorbed extracted from the FMR absorption curves. It should be noted that normalising for absorption only takes into account the quality of the bulk of the Py electrode. The interface is not necessarily the same between the samples and is a very significant factor in spin pumping. Having the two devices on the same chip, reduces this uncertainty.

The dominant contribution to the signal for both electrodes is the AMR voltage from the Py ferromagnet as in the previous LSP device. The relative weights of the $\mathrm{AMR^{x}}$ and $\mathrm{AMR^{z}}$ contributions depend on the exact orientation of the device with $\mathrm{AMR^{z}}$ only appearing if the device is not strictly perpendicular to the microwave current. Opposite signs of $\mathrm{AMR^{z}}$ of the two electrodes mean that the device was tilted in opposite directions between the two measurements. The $V_{\mathrm{ISHE}}$ component of around 2.5 $\mathrm{\mu V}$ was measured across the standard Pt electrode, identical to the LSP device in the previous section. Interestingly, a comparable magnitude of $V_{\mathrm{ISHE}}$ was also picked up by the discontinuous Pt electrode, which was not expected. Given that this electrode cannot create any genuine ISHE, we hypothesise the existence of another source of voltage that has the same $\mathrm{cos^{3}\mathnormal{\varphi}_{H}}$ dependence as the ISHE – we call it \emph{spurious} ISHE. It is most likely to originate in the Py film, given the number of other possible spurious effects related to FMR. To further investigate the origin and properties of this contribution we have performed further spin pumping measurements in bilayer systems, which are discussed in the next section.

Our findings highlight a further limitation of the LSP architecture used with conductive spin transport layers. Though the lateral architecture does not suffer from pinholes, which increase the coupling between the FM electrode and the detector electrode thus increasing the SREs weight in the SP voltage, the spurious voltages can still be dominant in a LSP device. Using a conductive material as a transport layer together with a conductive FM will inevitably lead to mixing of the pure ISHE signal and spurious ISHE generated in the FM. To avoid this, it is necessary to minimize SREs and/or use a material that exhibits efficient spin transport, but is not electrically conducting. Some promising candidates include organic-based magnet $\mathrm{V(TCNE)_x}$ (with $\mathrm{x\approx2}$) \cite{liu2018organic},\cite{liu2020spin} or inorganic magnetic oxides such as YIG \cite{manuilov2015spin}.

\section{Bilayer spin pumping}

\begin{figure}[t]
\begin{subfigure}{0.9\linewidth}
\caption{}
\includegraphics[width = \linewidth]{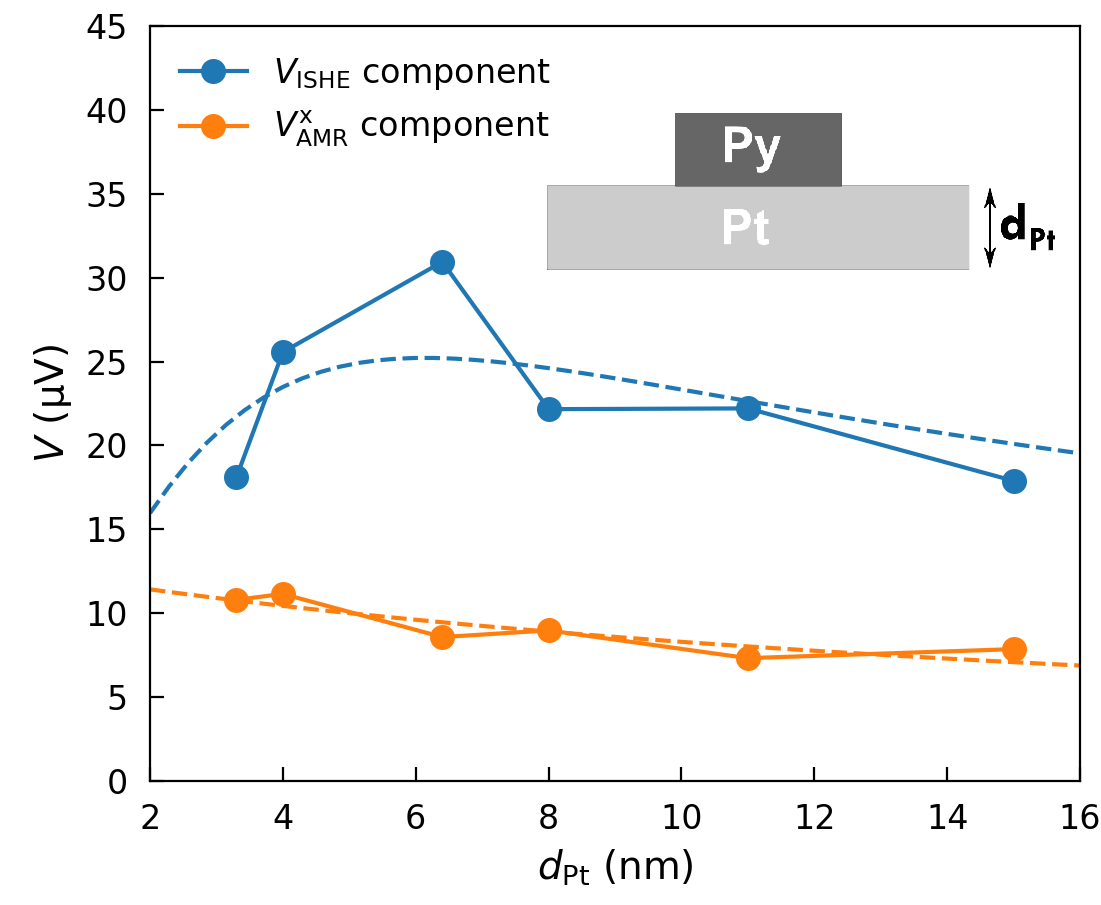}
\end{subfigure}
\begin{subfigure}{0.9\linewidth}
\caption{}
\includegraphics[width = \linewidth]{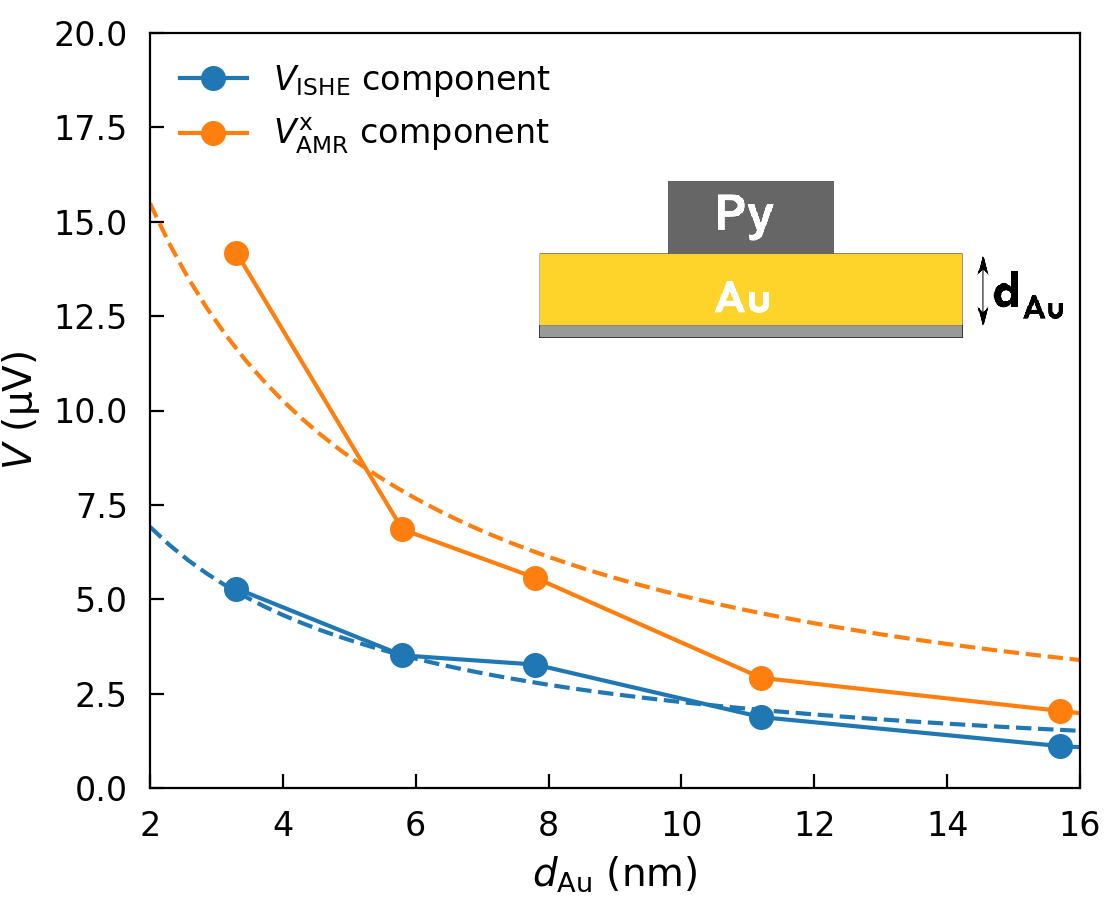}
\end{subfigure}
\caption{\label{fig:NM_thick_dep} $V_{\mathrm{ISHE}}$ and $V_{\mathrm{AMR}}^{\mathrm{x}}$ voltage contributions in Pt/Py (a) and Au/Py (b) bilayers as the non-magnet thickness was varied. Dotted lines show theoretical fits to the data.}
\end{figure}

Motivated to explore the spurious ISHE further, we have fabricated vertical bilayer spin pumping heterostructures where the Py layer is deposited directly on top of the non-magnetic layer, either Pt or Au. By measuring Pt/Py and Au/Py bilayers, we can separate the contribution from spurious ISHE generated in Py and compare it with ISHE in Pt. Measurements of the thickness dependence of each layer can provide further distinguishing characteristics.  Using thin films of Au, which has a smaller spin-Hall angle compared to Pt, allows us to reduce the genuine ISHE contribution and attribute the voltage signal to SRE and the spurious ISHE. The idea is similar to the discontinuous Pt electrode experiment, however, unlike the LSP architecture, bilayers allow for a direct contact between the non-magnet and the ferromagnet which provides a strong spin pumping efficiency in NM/FM structures. Moreover, it is relatively straightforward to account for the conductivity and electrical shunting of various layers in the bilayer architecture.

Figure \ref{fig:NM_thick_dep} shows non-magnetic layer thickness dependence of the $V_{\mathrm{AMR}}^{\mathrm{x}}$ and $V_{\mathrm{ISHE}}$ voltage components (the two dominant contributions) with the thickness of Py being constant at 25 nm. It can be seen from fig.~\ref{fig:NM_thick_dep}(a) that the $V_{\mathrm{ISHE}}$ component in Pt/Py bilayer has a typical non-monotonic thickness dependence in accordance with theoretical prediction for ISHE (eq.~\ref{eq:Vishe}). Initially, it increases as $\mathrm{tanh}\left(d_{\mathrm{Pt}}/2\lambda_{\mathrm{Pt}}\right)$ because more spin current is converted to charge current. It then saturates and starts decreasing on a larger length scale due to shunting of the signal by the conductance of the Pt layer as the thickness is increased. Numerical fit of eq.~\ref{eq:Vishe}, shown by the blue, dashed line, was obtained by keeping the functional form of thickness-dependent terms while pulling the remaining factors into one free parameter, giving a new scaling equation of the form $V_{\mathrm{ISHE}}=A\cdot \mathrm{tanh}\left(d_{\mathrm{Pt}}/2\lambda_{\mathrm{Pt}}\right)/\left(d_{\mathrm{Py}}\sigma_{\mathrm{Pt}}+d_{\mathrm{Py}}\sigma_{\mathrm{Py}}\right)$. We used fixed conductivity values of $\mathrm{\mathnormal{\sigma}_{Py}=1\times{10}^{6}\ Scm^{-1}}$ and $\mathrm{\mathnormal{\sigma}_{Pt}=1.3\times{10}^{6}\ Scm^{-1}}$ measured using reference samples with the van der Pauw method, while $A$ and $\lambda_{\mathrm{Pt}}$ were treated as the fitting parameters. The best fit value obtained for the spin diffusion length in Pt was $\mathrm{\mathnormal{\lambda}_{Pt}=1.9\ nm}$ which agrees with literature values \cite{tao2018self},\cite{roy2017estimating},\cite{zhang2013determination}. This provides strong evidence that the dominant signal in Pt/Py devices can certainly be attributed to the genuine ISHE. We stress that apart from the DC component of the ISHE measured in this experiment there also exists an AC component, however, it was disregarded in our experiments since it occurs in the gigahertz frequency range.

On the other hand, the $V_{\mathrm{ISHE}}$ component in Au/Py bilayer, shown in fig.~\ref{fig:NM_thick_dep}(b), has a significantly different profile. It is over six times smaller compared to Pt/Py devices and decreases monotonically. Given that $\mathrm{\mathnormal{\lambda}_{Au}\approx32\ nm}$ \cite{isasa2015temperature}, we would expect the ISHE in Au to show an initial increase over the range of thickness values probed in our experiment. Instead, the data can be modelled well with a thickness dependence expected for the AMR. Both the orange and blue dashed lines (as well as the orange dashed line from fig.~\ref{fig:NM_thick_dep}(a)) represent numerical fits of eq.~\ref{eq:sre-voltages}. It was obtained from the thickness dependent terms captured in the factor $f_{\mathrm{Py}}=d_{\mathrm{Py}}\sigma_{\mathrm{Py}}/\left(d_{\mathrm{Py}}\sigma_{\mathrm{Py}}+d_{\mathrm{Au}}\sigma_{\mathrm{Au}}\right)$, while the remaining terms were collected in one fitting parameter. We used $\mathrm{\mathnormal{\sigma}_{Au}=1.3\times{10}^{7}\ Scm^{-1}}$ together with the conductivity value for Py as before. Therefore, we conclude that the $V_{\mathrm{ISHE}}$ component in this case may be caused by a similar, spurious ISHE observed earlier in the discontinuous Pt electrode experiment. The similarity of its thickness dependence to that of rectified AMR suggests that it originates in the Py layer and that it might be yet another form of spin rectification.

\begin{figure}[t]
\begin{subfigure}{0.9\linewidth}
\caption{}
\includegraphics[width = \linewidth]{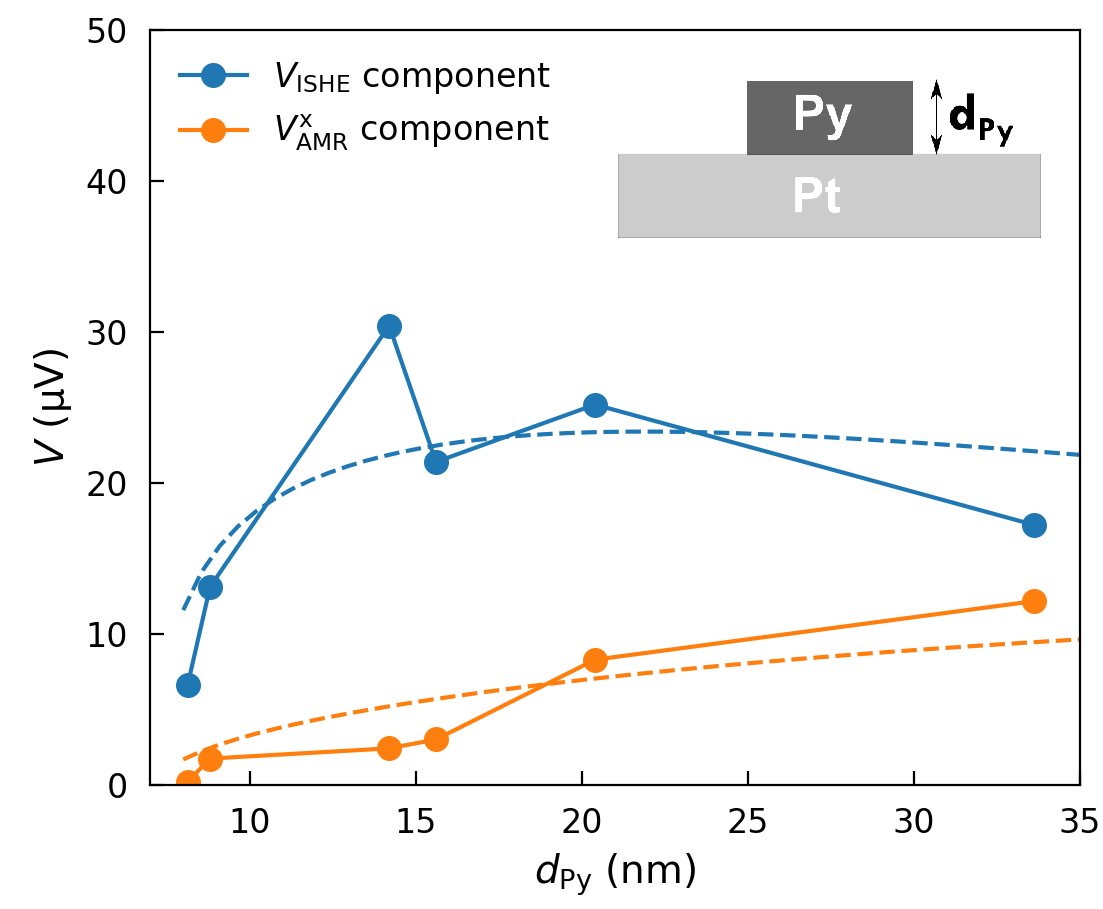}
\end{subfigure}
\begin{subfigure}{0.9\linewidth}
\caption{}
\includegraphics[width = \linewidth]{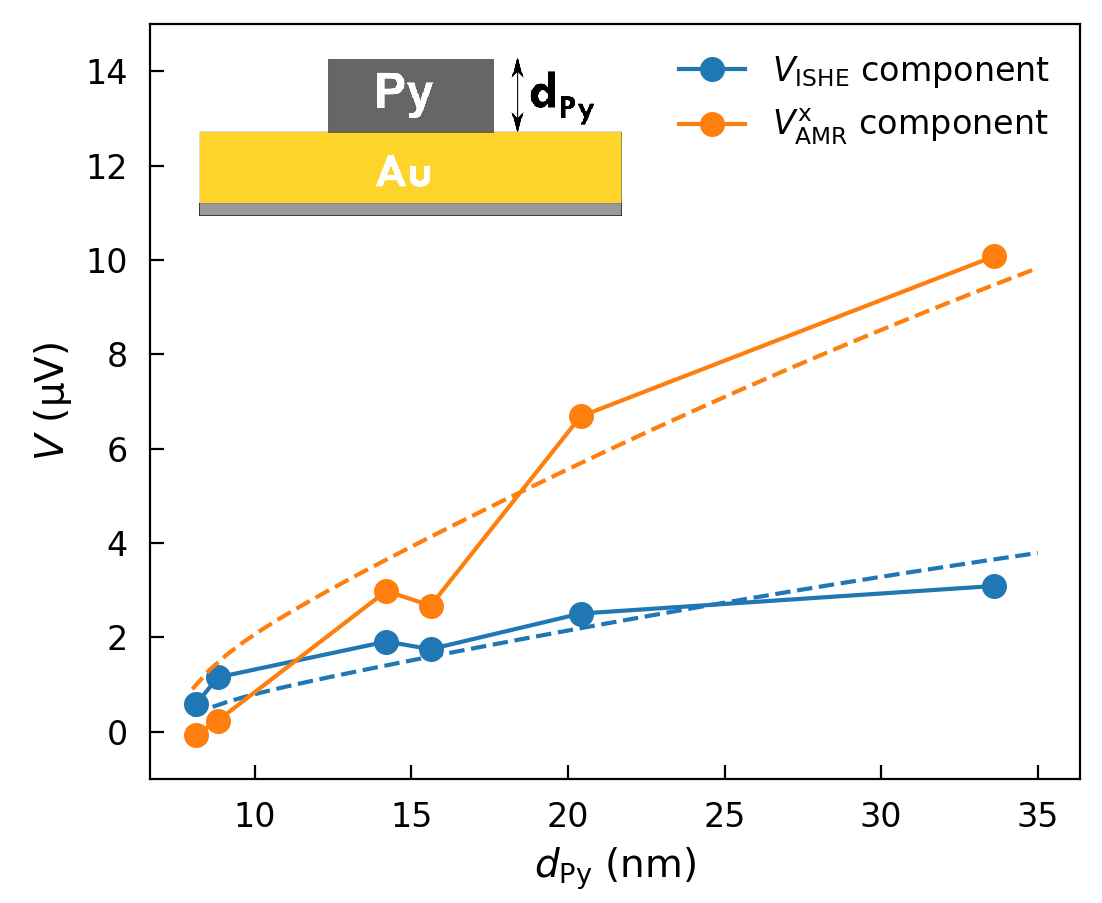}
\end{subfigure}
\caption{\label{fig:Py_thick_dep} $V_{\mathrm{ISHE}}$ and $V_{\mathrm{AMR}}^{\mathrm{x}}$ voltage contributions in Pt/Py (a) and Au/Py (b) bilayers as the Py thickness was varied. Dotted lines show theoretical fits to the data.}
\end{figure}

Further evidence distinguishing spurious ISHE from the genuine ISHE caused by spin pumping can be obtained by varying the thickness of the Py film. Figure \ref{fig:Py_thick_dep} shows the results of an experiment where the thickness of Pt and Au was kept constant at 7 nm and 5 nm respectively, while the thickness of Py was varied between 5 nm – 35 nm. For thicker Py films, the thickness dependence of the observed voltage components is captured by the factors $f_{\mathrm{Pt}}$ and $f_{\mathrm{Py}}$ in eqs.~\ref{eq:Vishe} and \ref{eq:sre-voltages}. The magnitude of SRE is expected to increase with $d_{\mathrm{Py}}$ and saturate at higher values, while the ISHE magnitude should decrease as the resistance of the shunting Py layer decreases. For thin Py films, modelling becomes more complicated since the Gilbert damping constant $\alpha$ increases rapidly with decreasing $d_{\mathrm{Py}}$ \cite{azevedo2011spin},\cite{tserkovnyak2002enhanced}, which leads to a decrease in both ISHE and SRE voltages. Furthermore, the equilibrium magnetisation $M_{\mathrm{0}}$ for ultra-thin Py films reduces, which leads to a further drop in voltage. Dashed lines in fig.~\ref{fig:Py_thick_dep} show numerical fits, which include the explicit modelling of $\alpha$ and $M_{\mathrm{0}}$ (more details in the appendix). While the $V_{\mathrm{ISHE}}$ component in fig.~\ref{fig:Py_thick_dep}(a) was fitted with eq.~\ref{eq:Vishe} which describes ISHE from spin pumping, both $V_{\mathrm{AMR}}^{\mathrm{x}}$ components as well as the $V_{\mathrm{ISHE}}$ component in fig.~\ref{fig:Py_thick_dep}(b) were fitted with eq.~\ref{eq:sre-voltages}, suitable for SRE. This is consistent with the previous observation and gives further support for the claim that the spurious ISHE originates in the Py layer.

The exact underlying mechanism of the spurious ISHE voltage, however, remains unclear. Thickness dependent measurements hint at the possibility that it is a form of spin rectification, which has not been accounted for in our analysis. Given the angular symmetry, it is also possible that it originates from self-induced ISHE in response to internal spin currents in Py. Further work is needed to identify phenomena that satisfy the required angular symmetry and agree with the measured thickness dependence.

\section{Conclusions}

The experiments shown in this paper demonstrate that spin rectification effects due to AMR and AHE are an important source of spurious voltages in lateral as well as vertical spin pumping devices with a conductive ferromagnet and need to be considered when interpreting the voltage signals generated. Simple checks for symmetries such as comparing measurements upon inversion of the magnetic field are insufficient to prove the observation of a genuine ISHE signal. However, it is possible to separate SREs from ISHE through a full in-plane angular dependence. Our measurements here suggest that the majority of the voltage signal observed in our previous LSP study on organic semiconductors \cite{wang2019long} should be interpreted as a rectified AMR signal rather than an ISHE signal. Even a small signal component that has the same angular symmetry as expected from a genuine ISHE signal that was detected both in LSP samples as well as in vertical bilayer systems may not be a genuine ISHE signal. Through careful control experiments we showed that it cannot be attributed to ISHE in Pt, but could rather be a yet unexplained effect originating in the Py layer. Our results emphasise that analysing spin pumping voltages is more complex than previously assumed due to the existence of spin rectification effects and likely other unknown sources of spurious voltages that have very similar symmetries and lineshapes compared to the ISHE. Our results point to a number of experimental design requirements for conducting successful spin pumping experiments with organic systems. These include: minimizing SRE through (a) careful optimisation of the thickness of the ferromagnet, or (b) the choice of a non-conducting ferromagnet such as YIG or organic ferromagnets like $\mathrm{V(TCNE)_x}$ (with $\mathrm{x\approx2}$), and/or (c) minimizing the conductivity of the spin transport layer.

\section{Acknowledgements}

The authors gratefully acknowledge funding from the European Research 
Council (ERC) through a Synergy Grant (Grant Agreement ID: 610115). D. 
Venkateshvaran acknowledges the Royal Society for funding in the form of 
a Royal Society University Research Fellowship (Royal Society Reference 
No. URF\textbackslash R1\textbackslash 201590). The authors express their thanks to Dr Radoslav 
Chakalov, Mr Roger Beadle and Mr Tom Sharp for technical support.

\appendix*
\section{}

\renewcommand{\thefigure}{A\arabic{figure}}
\setcounter{figure}{0}
\renewcommand{\theequation}{A\arabic{equation}}
\setcounter{equation}{0}

In order to obtain more accurate numerical fits of the Py thickness dependence experiments performed on vertical Pt/Py and Au/Py bilayers, the behaviour of the Gilbert damping parameter $\alpha$ has to be modelled. For thin ferromagnetic films it strongly depends on thickness, which influences both the spin pumping and spin rectification voltages through the components of the susceptibility tensor $\chi$ (eq.~\ref{eq:As_whole} in the main text). The Gilbert damping parameter can be written as \cite{tserkovnyak2002enhanced}
\begin{equation}
\alpha = \alpha_{\mathrm{0}} + \frac{\gamma \hbar g_{\mathrm{eff}}^{\mathrm{\uparrow\downarrow}}}{4\pi M_{\mathrm{0}}} \frac{1}{t_{\mathrm{FM}}},
\label{eq:alpha}
\end{equation}
where the first component $\alpha_{\mathrm{0}}$ is the intrinsic contribution and the second component is the additional damping due to spin pumping. Here, $\gamma$, $M_{\mathrm{0}}$, $g_{\mathrm{eff}}^{\mathrm{\uparrow\downarrow}}$ and $t_{\mathrm{FM}}$ are the gyromagnetic ratio, equilibrium magnetisation, effective spin-mixing conductance, and the thickness of the ferromagnetic layer. It is evident that the Gilbert damping parameter depends on the thickness of the ferromagnetic film in two ways: directly through the $1/t_{\mathrm{FM}}$ factor, and indirectly through the dependence on magnetisation. The latter dependence is especially important for very thin films in which the magnetisation decreases dramatically.

Magnetisation of thin films can be extracted from the FMR position, which is measured either from the microwave absorption or the induced voltage signal. For in-plane magnetic fields, the Kittel equation \cite{kittel1948on} describing the FMR condition takes the form
\begin{equation}
\left( \frac{\omega}{\gamma} \right)^{\mathrm{2}} = H_{\mathrm{FMR}} (H_{\mathrm{FMR}} + 4\pi M_{\mathrm{0}}),
\label{eq:kitel}
\end{equation}
and can be transformed to obtain $M_{\mathrm{0}}$ given $H_{\mathrm{FMR}}$ and the microwave frequency $\omega$. However, modelling magnetisation as a function of the ferromagnetic layer thickness is difficult, especially for an alloy such as Py which forms a multi-domain structure. Therefore, we use empirical modelling described below.

\begin{figure}[t]
\begin{subfigure}{0.9\linewidth}
\caption{}
\includegraphics[width = \linewidth]{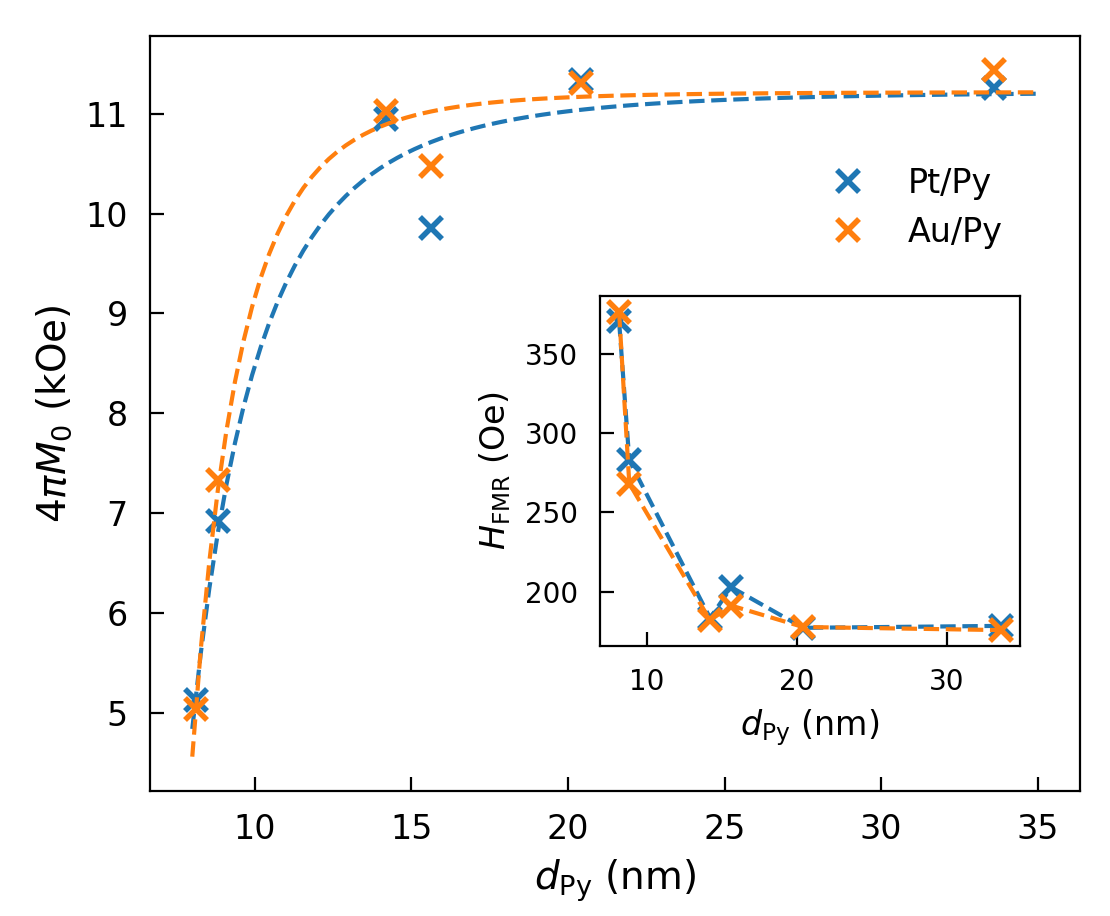}
\end{subfigure}
\begin{subfigure}{0.9\linewidth}
\caption{}
\includegraphics[width = \linewidth]{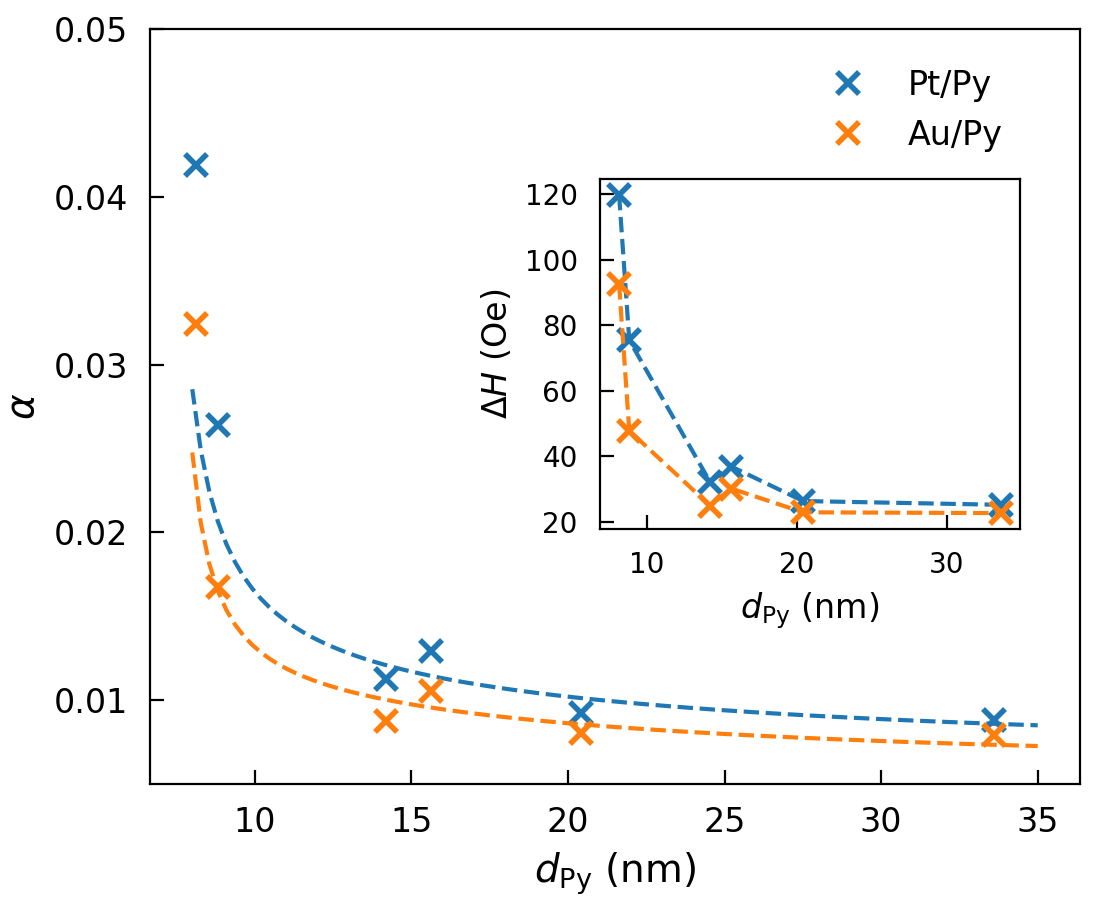}
\end{subfigure}
\caption{\label{fig:M0_modelling} Py thickness dependence of (a) the saturation magnetisation and (b) the Gilbert damping constant for two types of vertical bilayers with Pt (blue markers) and Au (orange markers) as the non-magnetic layer. Dashed lines show the numerical fits to the data.}
\end{figure}

The plot of the extracted magnetisation vs. thickness of the ferromagnetic layer for Au/Py and Pt/Py bilayers is shown in fig.~\ref{fig:M0_modelling}(a). It can be seen that $M_{\mathrm{0}}$ does not change with thickness for thicker films but decreases rapidly for very thin films on the order of 10nm. We propose to model this trend with an equation based on a modified power law
\begin{equation}
M_{\mathrm{0}}(t_{\mathrm{FM}}) = A - Bt_{\mathrm{FM}}^{\mathrm{-n}},
\label{eq:M0}
\end{equation}
where A,B and n are fitting parameters. Although such an empirical equation does not relate to a particular physical model, it provides an empirical, functional form to model magnetisation as a function of ferromagnetic film thickness, which is sufficient for the purpose of our experiments. The dashed lines in fig.~\ref{fig:M0_modelling}(a) are the resulting fits to the data – they demonstrate that this simple relation can model the magnetisation in the range of Py film thickness used in our experiments.

Figure \ref{fig:M0_modelling}(b) shows the values of the Gilbert damping parameter extracted from the FMR linewidth using the relation $\alpha=\Delta H\cdot\gamma/\omega$ \cite{gilbert2004a}. Using eq.~\ref{eq:alpha} together with the functional form of $M_{\mathrm{0}}$ extracted from fig.~\ref{fig:M0_modelling}(a), we produced numerical fits of the data (dashed lines in fig.~\ref{fig:M0_modelling}(b)) where $\alpha_{\mathrm{0}}$ and $g_{\mathrm{eff}}^{\mathrm{\uparrow\downarrow}}$ were treated as fitting parameters. We can see that $\alpha$ increases sharply for thin Py films, which is qualitatively consistent with eq.~\ref{eq:alpha}. However, the measured values for the thinnest films (8 nm) are significantly bigger than the numerical predictions. The parameters extracted in the process of numerical fitting are $\mathrm{\mathnormal{\alpha}_{0}=6.3\times{10}^{-3},\ \mathnormal{g}_{eff}^{\uparrow\downarrow}=4.63\times{10}^{15}\ cm^{-1}}$ for Pt/Py and $\mathrm{\mathnormal{\alpha}_{0}=5.5\times{10}^{-3},\ \mathnormal{g}_{eff}^{\uparrow\downarrow}=3.79\times{10}^{15}\ cm^{-1}}$ for Au/Py bilayers. While $\alpha_{\mathrm{0}}$ in both cases is comparable to the literature values \cite{azevedo2011spin}, the effective spin-mixing conductance values are a few times bigger than previously reported. Consequently, the Gilbert damping constant of around 0.04 measured for 8 nm thick Py films is a few times bigger than reported in literature \cite{azevedo2011spin}. These discrepancies are suggestive of an extra broadening mechanism that increases damping in our experiments. It is likely to be caused by a non-uniform growth of the film at such small thickness values which results in a rough surface and influences the formation of magnetic domains.

The above modelling of the Gilbert damping constant using eq.~\ref{eq:alpha} captures the general trend, however, it does not take into account the non-uniform growth of the ferromagnetic film, hence failing for very thin films. While the full modelling is beyond the scope of this paper, we note that this simple model fits the data reasonably well and therefore can be used to improve the numerical fits of Py thickness dependence of the voltage signal reported in the main text.

\bibliography{bibliography}

\end{document}